\documentclass[prd,showpacs,showkeys,nofootinbib,eqsecnum,floatfix,fleqn,
               preprint,12pt,tightenlines]{revtex4} 

\usepackage{amsmath,amssymb,revsymb,graphicx,dcolumn}

\newcommand{\textwidthPreprintTwocolumn}{0.85\textwidth} 

\newcommand{\beq}{\begin{equation}}
\newcommand{\eeq}{\end{equation}}
\newcommand{\beqa}{\begin{eqnarray}}
\newcommand{\eeqa}{\end{eqnarray}}
\newcommand{\bsubeqs}{\begin{subequations}}
\newcommand{\esubeqs}{\end{subequations}}

\newcommand{\half}{{\textstyle \frac{1}{2}}}    

\newcommand{\lambdaOneTwo}{\lambda_{12}}  
\newcommand{\lambdaTwoOne}{\lambda_{21}}

\begin{document}

\noindent
Phys. Rev. D 82, 083006 (2010)
\hfill
arXiv:1001.1939\newline\vspace*{2mm}
\title{Effective cosmological constant from TeV--scale physics\vspace*{5mm}}
\author{F.R. Klinkhamer}
\email{frans.klinkhamer@kit.edu}
\affiliation{\mbox{Institute for Theoretical Physics, University of Karlsruhe,}
Karlsruhe Institute of Technology, 76128 Karlsruhe, Germany\\}

\begin{abstract}
\vspace*{2.5mm}\noindent
It has been suggested previously that
the observed cosmological constant $\Lambda$
corresponds to the remnant vacuum energy density
of dynamical processes taking place at a cosmic age set by
the mass scale $M \sim E_\text{ew}$ of ultramassive particles
with electroweak interactions. Here, a simple modification
of the nondissipative dynamic equations of $q$--theory is presented,
which produces a remnant vacuum energy density
(effective cosmological constant) of the correct order of magnitude.
Combined with the observed value of $\Lambda$, a first estimate of
the required value of the energy scale $E_\text{ew}$ ranges from
3 to 9 TeV, depending on the number of species of ultramassive particles
and assuming a dissipative coupling constant of order unity.
If correct, this estimate implies the existence of new
TeV--scale physics beyond the standard model.
\end{abstract}

\maketitle

\section{Introduction}
\label{sec:Introduction}

It has been argued by Arkani-Hamed \emph{et al.}~\cite{ArkaniHamed-etal2000}
that two fundamental energy scales,
the electroweak scale $E_\text{ew} \sim 1\;\text{TeV}$
and the gravitational scale $E_\text{Planck}\sim 10^{15}\;\text{TeV}$,
suffice to explain the triple  cosmic coincidence puzzle:
why are the orders of magnitude of the energy densities of vacuum, matter,
and radiation approximately the same in the present Universe?
For this explanation to work,  the parametric form of the effective
cosmological constant (remnant vacuum energy density) must be
\beq\label{eq:rhoV-EW-formula}
\Lambda \equiv \rho_\text{V,\,remnant}
\sim \big( (E_\text{ew})^{2}\, /E_\text{Planck} \big)^{4}
       \sim (10^{-3}\;\text{eV})^4 \,.
\eeq
If true,  formula \eqref{eq:rhoV-EW-formula} would be a remarkable
explanation of the measured value from observational cosmology,
which appears to be of order
$10^{-29}\;\text{g}\,\text{cm}^{-3}$ $\sim$ $10^{-11}\;\text{eV}^4$
(setting $\hbar=c=1$ and referring to, e.g.,
Refs.~\cite{Riess-etal1998,Perlmutter-etal1998,Komatsu2008}
and other references therein).\footnote{There have, of course,
been many other explanations of the smallness of $\Lambda$
by an appropriate ratio of energy scales
(see, e.g., Sec.~X of Ref.~\cite{RatraPeebles1988}), but the
relation \eqref{eq:rhoV-EW-formula} is special as it carries
the ingredients to naturally give the correct orders of magnitude
for the present matter and radiation energy
densities~\cite{ArkaniHamed-etal2000}.}
However, \eqref{eq:rhoV-EW-formula}  was not derived convincingly in
Ref.~\cite{ArkaniHamed-etal2000}, as an unknown adjustment mechanism
needed to be invoked.

Subsequently, Volovik and the present author
realized~\cite{KV2009-electroweak} that,
in the framework of \mbox{$q$--theory,} there is the possibility
of generating a vacuum energy density precisely of the
form \eqref{eq:rhoV-EW-formula}.
Here, $q$--theory is a particular approach~\cite{KV2009-CCP1}
to solving the first cosmological constant problem (CCP1):
why is $|\Lambda| \ll (E_\text{Planck})^4\,$?
The original references on the statics and dynamics of
$q$--theory are~\cite{KV2008-statics}
and~\cite{KV2008-dynamics}, respectively.
The second cosmological constant problem (CCP2) is
the question addressed here, namely, the actual order of
magnitude of $\Lambda$, if indeed nonzero.

The positive remnant vacuum energy density obtained in
Ref.~\cite{KV2009-electroweak} relied crucially on Eq.~(4.1) of that
article. That particular equation was taken to
describe the quantum-dissipative
effects of the vacuum energy density, but was, in the end,
purely hypothetical
and disconnected from the previous $q$--theory discussion.

The question arises if it is at all possible to modify the previous
$q$--theory equations~\cite{KV2008-dynamics}
in such a way as to effectively recover
the results of Ref.~\cite{KV2009-electroweak}.
The answer is that this is indeed possible, even though the
required modifications are quite subtle.
The scope of the present article is restricted to finding these
appropriate \emph{phenomenological} equations,
rather than establishing the relevant
microscopic processes of the underlying theory.

In a way, it can be said that this whole article
is about the proportionality constant implicit in \eqref{eq:rhoV-EW-formula}
and that, within the framework of $q$--theory,
the article gives an existence proof
for a set of dynamical equations which produces a proportionality
constant of order unity.

As the previous discussion makes clear,
the present article is a direct follow-up of
Ref.~\cite{KV2009-electroweak}, to which the reader is referred
for the original motivation and detailed analysis.
This article is, by necessity, rather technical and
it may be helpful to give the reader a road map.
The material of this article is organized as follows:
\begin{description}
\item[\;\;\; Track~1:]
Secs.~\ref{sec:Introduction}, \ref{sec:Theoretical-framework},
\ref{subsec:Basic-idea}, \ref{subsec:Modified-ODEs},
and \ref{sec:Discussion};
\item[\;\;\; Track~2:]
Secs.~\ref{subsec:Ansaetze},
\ref{subsec:Dimensionless-ODEs}, and \ref{sec:Numerical-solution};
\item[\;\;\; Track~3:]
Sec.~\ref{subsec:Additional-remarks} and Appendix~\ref{sec:appendixA}.
\end{description}
The basic idea and main results are presented in Track~1,
the dimensionless differential equations
and their numerical solution in Track~2,
and a more detailed discussion and further refinement in Track~3
(the most realistic calculations are
shown in the very last two figures and the very last table
of Appendix~\ref{sec:appendixA}).
In a first reading, it is possible to follow Track~1 and to add
the other Tracks later.

\section{Theoretical framework}
\label{sec:Theoretical-framework}

This section reviews the main ingredients of the type of
theory considered in this article
(see Refs.~\cite{KV2008-statics,KV2008-dynamics} for details).
The particular $q$--theory realization used
involves the so-called 4-form
field strength~\cite{DuffNieuwenhuizen1980,Aurilia-etal1980}.
Very briefly, the theory is defined over a four-dimensional Lorentzian
spacetime manifold and employs a 4-form field strength $F$
derived from a 3-form gauge field $A$.
The corresponding rank-four tensor can always be written as
\beq
F_{\alpha\beta\gamma\delta}(x) =
q(x)\,\sqrt{-g(x)}\, \epsilon_{\alpha\beta\gamma\delta}(x)\,,
\eeq
with the Levi--Civita tensor density $\epsilon_{\alpha\beta\gamma\delta}(x)$,
the determinant of the metric $g(x)\equiv \det g_{\alpha\beta}(x)$,
and the scalar field $q(x)$. The crucial point is that
this scalar field $q(x)$ is \emph{nonfundamental},
being built from the metric field $g_{\alpha\beta}(x)$ and
the 3-form gauge field $A(x)$, as will become clear shortly.
This 3-form gauge field $A(x)$ is considered to be one
of the fields which characterize the quantum vacuum at the
fundamental microscopic level~\cite{KV2008-statics}.

The macroscopic effective action of the relevant high-energy
fields (here, $A$) and the low-energy fields (here, $g$ and $\psi$)
is taken to be of the following form~\cite{KV2008-dynamics}:
\bsubeqs\label{eq:action-Seff-q2def-Fdef}
\beqa
S_\text{eff}[A, g,\psi]&=&
\int_{\mathbb{R}^4} \,d^4x\, \sqrt{-g}\,\Big( K(q)\,R[g]
+ \epsilon(q)+\mathcal{L}^{M}[\psi,g] \Big)\,,
\label{eq:action-Seff}
\\[2mm]
F_{\alpha\beta\gamma\delta}&=& \nabla_{[\alpha}A_{\beta\gamma\delta]}\,,
\label{eq:Fdef}\\[2mm]
q^2 &\equiv& - \frac{1}{24}\, F_{\alpha\beta\gamma\delta}\,
                              F^{\alpha\beta\gamma\delta}\,,
\label{eq:q2def}
\eeqa
\esubeqs
where the effective gravitational coupling parameter $K$
is allowed to depend on $q$,
$R[g]$ is the Ricci curvature scalar obtained from the metric
$g_{\alpha\beta}$, $\nabla_\alpha$ denotes the standard covariant derivative,
and the square bracket around spacetime indices stands for
complete antisymmetrization. The energy density $\epsilon(q)$
is assumed to be a
generic function of $q$, that is, a function different
from the simple quadratic $\half\,q^2$ corresponding to a
Maxwell-type theory~\cite{DuffNieuwenhuizen1980,Aurilia-etal1980}.
The field $\psi$ in \eqref{eq:action-Seff} stands for a generic low-energy matter
field with a scalar Lagrange
density $\mathcal{L}^{M}[\psi,g]$, which,
for simplicity, is assumed to be without explicit $q$--field dependence
(the dependence on the metric arises from the covariant derivatives).

Remark that the effective action \eqref{eq:action-Seff} corresponds to a
Brans--Dicke-type action \cite{BransDicke1961},
but without kinetic term for the (nonfundamental) scalar $q$.
For spacetime-independent $q$ [that is, $q(x)=\overline{q}=\text{const}$],
the effective action \eqref{eq:action-Seff} corresponds to
the one of standard general relativity with a cosmological constant
$\Lambda=\epsilon(\overline{q})+\Lambda^{M}$,
where $\Lambda^{M}$ refers to contributions to $\Lambda$
from the matter Lagrange density $\mathcal{L}^{M}$.

By taking variations of $A_{\alpha\beta\gamma}(x)$
and $g_{\alpha\beta}(x)$ in the effective action  \eqref{eq:action-Seff},
generalized Maxwell and Einstein equations can be derived.
The generalized Maxwell equation can be solved explicitly
and the solution depends on a constant of integration $\mu$.
With the solution of the generalized
Maxwell equation, the generalized Einstein equation reduces to the following
field equation:
\begin{eqnarray}
\hspace*{-5mm}
2K\,\big( R_{\alpha\beta}-g_{\alpha\beta}\,R/2 \big)&=&
-2\,\big(  \nabla_\alpha\nabla_\beta - g_{\alpha\beta}\, \square\big)\, K(q)
+\rho_{V}(q)\, g_{\alpha\beta} - T_{\alpha\beta}^{M} \,,
\label{eq:reduced-gen-Einstein}
\end{eqnarray}
where the combination
\beq
\rho_{V}(q) \equiv  \epsilon(q)-  \mu\, q
\label{eq:rhoV-def}
\eeq
plays the role of the \emph{gravitating} vacuum energy density
rather than the single term $\epsilon(q)$
appearing in the effective action \eqref{eq:action-Seff}.
Furthermore, there is an equation remaining from the particular
solution of the generalized Maxwell equation,
which reads
\beq\label{eq:reduced-gen-Maxwell}
\frac{d\rho_{V}}{d q} + R\,\frac{d K}{d q} =0\,.
\eeq

The final equations
\eqref{eq:reduced-gen-Einstein}--\eqref{eq:reduced-gen-Maxwell}
can be specialized to the case of a
spatially flat Friedmann--Robertson--Walker universe. The resulting
cosmological equations have been studied in
Refs.~\cite{KV2009-electroweak,KV2008-dynamics} and it is the aim of
the present article to find a modification of them
which allows for the generation of a nonvanishing remnant
vacuum energy density.

\section{Vacuum dynamics in a flat FRW universe}
\label{sec:Vacuum-dynamics}

\subsection{Basic idea}
\label{subsec:Basic-idea}
\vspace*{-1mm}

Following Ref.~\cite{ArkaniHamed-etal2000},
assume the existence of ultramassive unstable particles
(here, called `type 1a') with masses $M$ of
order $E_\text{ew} \sim 1\;\text{TeV}$ and electroweak interactions.
Consider a spatially flat Friedmann--Robertson--Walker (FRW) universe
and assume the type--1a particles to be effectively
in thermal equilibrium at early enough times.
Then, the masses of these particles
start to affect the Hubble expansion rate $H(t)$
when the temperature drops to $T \sim E_\text{ew}$,
corresponding to a cosmic age $t$ of order
\beq\label{eq:t_ew-def}
t_\text{ew}  \equiv  E_\text{Planck}/(E_\text{ew})^2 \,,
\eeq
in terms of the reduced Planck energy,
\beq\label{eq:EPlanck-def}
E_\text{Planck}
\equiv
\sqrt{1/(8\pi G_{N})}
\approx
2.44\times 10^{18}\:\text{GeV}.
\eeq
Note that the definition \eqref{eq:t_ew-def}
is motivated by the standard Friedmann equation
\mbox{$H^2 \sim \rho/(E_\text{Planck})^2$}
with $H \sim 1/t_\text{ew}$ and $\rho \sim T^4 \sim (E_\text{ew})^4$.
In the following, the Friedmann equation will be modified,
but the order of magnitude \eqref{eq:t_ew-def} remains relevant.
Throughout this article, natural units are used with $\hbar=c=1$.

Compared to the case of having only ultrarelativistic particles
(these lighter particles are called `type 1b' and can be thought to
have masses of order $M/10$),
the change of the expansion rate from the
ultramassive type--1a particles can be modeled by
a nonzero function $\kappa_{M1}(t/t_\text{ew})$.
In fact, this function $\kappa_{M1}$ can be written
in terms of the standard equation-of-state (EOS) parameter
$w_{M1}\equiv P_{M1}/\rho_{M1}$
as follows:
\begin{equation}
\kappa_{M1} \equiv 1 - 3\, w_{M1}\,,
\label{eq:kappaM1}
\end{equation}
which vanishes for ultrarelativistic particles ($w_{M1}=1/3$)
and equals unity for pressureless nonrelativistic particles ($w_{M1}=0$).
By taking different unstable type--1 particles  ($a,\,b,\,c,\,\ldots$)
it is possible to obtain an effective
EOS function $\kappa_{M1}(t/t_\text{ew})$
which peaks at $t=t_\text{ew}$. Here, however, a particular form
of $\kappa_{M1}$ will simply be assumed.

The main conditions on this assumed EOS function $\kappa_{M1}$
are that it
peaks at $t=t_\text{ew}$ and is nonzero only in a finite range
around the maximum (this last condition is not essential but simplifies
the discussion). Specifically, the conditions are taken to be:
\bsubeqs\label{eq:kappaM1-conditions}
\beqa
\kappa_{M1}(t/t_\text{ew}) &<&  \kappa_{M1}(1)
\quad
\text{for}\;\; t \ne t_\text{ew} \,,
\label{eq:kappaM1-conditions-peak}
\\[.6mm]
\kappa_{M1}(t/t_\text{ew}) &\ne& 0
\quad\hspace*{10.5mm}
\text{for}\;\; t \in \big(t_\text{start},t_\text{end}\big) \,,
\label{eq:kappaM1-conditions-nonzero}
\\[.5mm]
\kappa_{M1}(t/t_\text{ew}) &=& 0
\quad\hspace*{10.5mm}
\text{for}\;\; t \notin \big(t_\text{start},t_\text{end}\big) \,,
\label{eq:kappaM1-conditions-zero}
\eeqa
\esubeqs
with $0 < t_\text{start} < t_\text{ew}$
and $t_\text{ew} < t_\text{end} < \infty$,
having set $t=0$ for the big bang where $H(t)$ diverges.
The physical picture corresponding to \eqref{eq:kappaM1-conditions}
is that ultramassive type--1a particles
are dominant at $t \sim t_\text{ew}$, but, then, decay into lighter
type--1b particles which are still ultrarelativistic for $t$
not very much larger than $t_\text{ew}$.
The main goal is to study the effects of this prescribed EOS
function $\kappa_{M1}$, relegating the discussion of a more realistic
EOS function to App.~\ref{sec:appendixA}.

A flat FRW universe containing only type--1 particles with a prescribed
EOS parameter \eqref{eq:kappaM1}--\eqref{eq:kappaM1-conditions}
has a standard radiation-dominated
Hubble expansion rate $H(t)=(1/2)\,t^{-1}$ for
$t <t_\text{start}$ and $t>t_\text{end}$.
The expansion rate is changed,
$H(t)\ne(1/2)\,t^{-1}$,
for times $t$ between $t_\text{start}$  and $t_\text{end}$.
The question, now, is what happens if this FRW universe
also has a dynamical vacuum-energy-density component.

For the theory outlined in Sec.~\ref{sec:Theoretical-framework},
two results were obtained in Ref.~\cite{KV2009-electroweak}.
First, it was shown that there is an exact solution having
$\rho_{V}(t)=0$ in the radiation-dominated phase
with $\kappa_{M1}(t)=0$.
Second, it was shown that the changed Hubble expansion
from $\kappa_{M1}(t)\ne 0$
\emph{kicks} $\rho_{V}(t)$ away from zero.
Specifically, the following behavior was
established~\cite{KV2009-electroweak}:
\begin{equation}
\rho_{V}(t) \sim \kappa_{M1}^2(t)\, H(t)^4 \,,
\label{eq:rhoV-kick}
\end{equation}
which vanishes asymptotically as $\kappa_{M1}$ drops to zero and the
standard radiation-dominated expansion of the model universe
resumes. At the moment of the kick, $t \sim t_\text{ew}$,
the vacuum energy density \eqref{eq:rhoV-kick} is of order
$(t_\text{ew})^{-4}
\sim \big((E_\text{ew})^2/E_\text{Planck}\big)^4
\sim \big(E_\text{ew}/E_\text{Planck}\big)^4\,(E_\text{ew})^4$,
which is negligible compared to the
matter energy density $\rho_{M1} \sim(E_\text{ew})^4$.
The vacuum energy density $\rho_{V}(t)$, therefore, just responds to
(is being kicked by) the Hubble expansion
and does not affect the expansion substantially.

The result \eqref{eq:rhoV-kick} has been obtained
from the simplest time-reversible
(nondissipative) version of $q$--theory, with field equations
given by \eqref{eq:reduced-gen-Einstein}--\eqref{eq:reduced-gen-Maxwell}.
It has been argued that quantum-dissipative effects
(e.g., because of particle production in an expanding
universe~\cite{ZeldovichStarobinsky1977,BirrellDavies1982}) may result
in a freezing of the previous result \eqref{eq:rhoV-kick} to a constant
nonzero value.

As explained in Sec.~\ref{sec:Introduction}, the aim
of this article is to find a
suitable modification of the ``classical'' $q$--theory equations,
which produces a finite remnant vacuum energy density.
In the approach followed here,  there are three changes:
\begin{enumerate}
\item
The matter energy-conservation equation is modified to include
appropriate particle-production effects operating at a cosmic
age $t \sim t_\text{ew}$.
\item
The reduced Maxwell equation is modified, so as to match
the standard Einstein equation of an FRW universe
with a nonzero effective cosmological constant at later times.
\item
Different particle species are considered with ultramassive
type--1a particles first decaying into
lighter type--1b particles, which, in turn,
decay into massless type--2 particles.
\end{enumerate}
The first two modifications are essential for the generation of
a nonvanishing remnant vacuum energy density.
(The first modification has already been discussed
in general terms in Sec.~IV of Ref.~\cite{KV2009-electroweak}).
The third modification allows for
a possibly more realistic scenario, with type--1 particles corresponding
to new TeV-scale physics and type--2 particles corresponding to
the standard model of elementary particle physics
(see also App.~\ref{sec:appendixA}).

The particle-production effects of point 1 above will be
controlled by an effective coupling constant $\zeta>0$
and a particular type of dissipation function
$\gamma(t/t_\text{ew};\zeta)\geq 0$.
The reason for calling $\gamma$ a ``dissipation'' function
will become clear in Sec.~\ref{subsec:Additional-remarks}.
The main conditions on this function are as follows:
\bsubeqs\label{eq:gamma-conditions}
\beqa
\gamma(0;\zeta) &=& 1\,,
\label{eq:gamma-conditions-bc0}
\\[1mm]
\forall \, t\geq t_\text{freeze}:\;\;  \gamma(t/t_\text{ew};\zeta) &=& 0 \,,
\label{eq:gamma-conditions-bc-larget}
\\[1mm]
\lim_{\zeta\to 0}\; (t_\text{freeze})^{-1} &=& 0\,,
\label{eq:gamma-conditions-tfreezelimit}
\eeqa
\esubeqs
with a particular time $t_\text{freeze}$ that is of order $t_\text{ew}$
for $\zeta \sim 1$ and approaches infinity for $\zeta \to 0$.

The coupling constant $\zeta$ and the corresponding function $\gamma$
are purely phenomenological.
The vanishing of $\gamma$ for large enough times will be seen to
have two effects: first, to freeze the ``classical'' value \eqref{eq:rhoV-kick}
and, second, to switch to a standard FRW expansion with relativistic matter
and a tiny value \eqref{eq:rhoV-EW-formula}
for the remnant vacuum energy density.
Further discussion of $\zeta$ and $\gamma$ will appear in
the next subsection, after the differential equations
have been presented.

\subsection{Modified ODEs}
\label{subsec:Modified-ODEs}

For a spatially flat FRW universe and the 4-form realization of
$q$--theory with a variable gravitational coupling parameter $K(q)$,
the cosmological differential equations have been derived
in Ref.~\cite{KV2008-dynamics} and were already mentioned
in Sec.~\ref{sec:Theoretical-framework}. The basic idea of the
proposed modification of these ordinary differential equations (ODEs)
has been discussed in the previous subsection. Specifically,
the modified ODEs are given by:
\bsubeqs\label{eq:modifiedODEs}
\beqa
\hspace*{-12mm}
6\, \frac{d K}{d q} \left(\frac{d H}{d t} +2H^2 \right)
&=&  
\big[\gamma(t/t_\text{ew})\big]\,\frac{d\rho_{V}}{d q}
+ \big[ 1- \gamma(t/t_\text{ew})  \big]\,\frac{1}{K} \,\frac{d K}{d q}\:2\rho_{V}\,,
\label{eq:modifiedODEs-Hdot}\\[2mm]
\hspace*{-12mm}
\frac{d\rho_{M1}}{d t}+
\big[4-\kappa_{M1}(t/t_\text{ew})\big]\,H\,\rho_{M1}
&=&  
- \frac{\zeta}{\gamma(t/t_\text{ew})}\:
  q \frac{d}{d t}\left(\frac{d\rho_{V}}{d q}\right)
- \frac{\lambdaOneTwo}{t_\text{ew}}\, \big[ 1- \gamma(t/t_\text{ew}) \big]\,\rho_{M1}  \,,
\label{eq:modifiedODEs-M1dot}\\[2mm]
\hspace*{-12mm}
\frac{d\rho_{M2}}{d t}+4\,H\,\rho_{M2}
&=&  
+\frac{\lambdaOneTwo}{t_\text{ew}}\, \big[ 1- \gamma(t/t_\text{ew}) \big]\,\rho_{M1} \,,
\label{eq:modifiedODEs-M2dot}\\[2mm]
\hspace*{-12mm}
6\,\left(
H\,\frac{d K}{d q}\,\frac{d q}{d t}+ K\,H^2 \right)
&=&  
\rho_{V}+\rho_{M1}+\rho_{M2} \,,
\label{eq:modifiedODEs-qdot}
\eeqa
\esubeqs
where only the arguments of the functions
$\kappa_{M1}$ and $\gamma$ have been
shown explicitly. The four equations in \eqref{eq:modifiedODEs},
going from the top to the bottom,
can be recognized as modified versions of the reduced Maxwell equation
\eqref{eq:reduced-gen-Maxwell}, the two matter energy-conservation equations,
and the Friedmann equation [the standard Friedmann equation is recovered for
$K=K(q_0)=1/(16\pi G_\text{N})=\text{const}$].
For $\zeta=0$ and $\gamma(t)=1$, the ODEs \eqref{eq:modifiedODEs}
correspond to Eqs.~(4.12abc) of Ref.~\cite{KV2008-dynamics},
supplemented by an equation for the adiabatic evolution of $\rho_{M2}$.

The modified ODEs for $\zeta > 0$ have a dependence on an external
time scale, here taken to be $t_\text{ew}$ from \eqref{eq:t_ew-def}.
The $t_\text{ew}$  dependence enters implicitly through the EOS function $\kappa_{M1}\geq 0$ and the dissipation function $\gamma\geq 0$
discussed in the previous subsection and
explicitly through the matter energy-exchange terms proportional to
$\lambdaOneTwo/t_\text{ew}$.
Note that the dimension of the $q$--field in Eqs.~\eqref{eq:modifiedODEs}
is irrelevant, which concords with the fact that
this $q$--field may be realized in different
ways~\cite{KV2008-statics}.

The particular modification \eqref{eq:modifiedODEs} of the
cosmological ODEs from classical $q$--theory has two main ingredients:
first, the function $\gamma(t/t_\text{ew})$ with characteristics
\eqref{eq:gamma-conditions} and, second, the presence of a finite coupling
constant $\zeta$,
\beq
\zeta = \text{O}(1)\,,
\eeq
which enters directly on the right-hand side
of \eqref{eq:modifiedODEs-M1dot} and
indirectly via condition \eqref{eq:gamma-conditions-tfreezelimit} for
the dissipation function $\gamma$.
A finite value for the remnant vacuum energy density from the
dynamical ODEs \eqref{eq:modifiedODEs} requires both
$\zeta> 0$ and $\zeta <\infty$, as will be explained in
Sec.~\ref{subsec:Additional-remarks}.

The rather simple structure of \eqref{eq:modifiedODEs},
combined with conditions \eqref{eq:kappaM1-conditions} and
\eqref{eq:gamma-conditions},
will be seen to allow for the generation of a nonzero
remnant vacuum energy density.\footnote{A somewhat more general
modification of the ODEs has $2\rho_{V}$ in the last term
on the right-hand side of \eqref{eq:modifiedODEs-Hdot} replaced by
$[2\rho_{V}+\kappa_{M1}(t/t_\text{ew})\,\rho_{M1}/2]$,
but, for the case considered in this section and
Sec.~\ref{sec:Numerical-solution},
the results are essentially unchanged,
because $\kappa_{M1}(t/t_\text{ew})=0$ for  $t \geq t_\text{end} > t_\text{ew}$
according to \eqref{eq:kappaM1-conditions}.\label{ftn:changed-mod-Maxwell-eq}}
A detailed discussion of the modified ODEs is postponed until
Sec.~\ref{subsec:Additional-remarks}, after
these equations have been established in dimensionless form.

Before embarking on this technical enterprise,
it may be useful to recapitulate the basic assumptions.
The first assumption is
the existence of a particular type of vacuum variable $q(x)$, namely, a variable
which corresponds to
a conserved relativistic quantity $q_0$ in flat Minkowski spacetime.
Such a variable $q(x)$ provides a possible solution of the main
cosmological constant problem (CCP1)
by explaining why $\Lambda/E_\text{Planck}^4$
is naturally zero in the equilibrium state.\footnote{The
$q$--theory approach to CCP1 provides only a \emph{possible} solution,
because it is not known for sure that the underlying microscopic
theory \emph{does} contain an appropriate $q$--type field.
In addition, there remain other equally fundamental
(perhaps related) questions, such as the nature of gravity
and the origin of spacetime.
The goal of the present article is relatively modest:
to explore, in the framework of $q$--theory, a possible connection
between the observed value of the effective cosmological constant
and new TeV--scale physics.} This vacuum variable $q(x)$ is taken
to have an effective action of the form of \eqref{eq:action-Seff-q2def-Fdef},
where, in particular, the gravitational coupling constant $K$ may carry a
dependence on $q$.

The second assumption is that the field equations from the effective action
\eqref{eq:action-Seff-q2def-Fdef}, specialized to a spatially flat
FRW universe,
are modified by the introduction of terms involving the coupling constant
$\zeta$. The crucial term is the first one on the right-hand side of
\eqref{eq:modifiedODEs-M1dot}, whose physical motivation is that it
reproduces the dissipative behavior suggested in
Ref.~\cite{KV2009-electroweak} (this behavior is
analogous to that of bulk viscosity in compressible material
fluids~\cite{LandauLifshitz-Fluid-mechanics}).
The coupling constant $\zeta$ and the corresponding function
$\gamma$ are purely phenomenological.
As mentioned in Sec.~\ref{sec:Introduction}, ultimately
$\zeta$ and $\gamma$ (or appropriate generalizations)
need to be derived from the underlying microscopic theory,
but that task lies outside the scope of the present article.

Clearly, the first assumption is better motivated than the second.
But the second assumption may (or may not) gain in credibility
depending on the success (or not) of producing a reasonable remnant
vacuum energy density and predicting new TeV--scale physics.

\subsection{Ans\"{a}tze and dimensionless variables}
\label{subsec:Ansaetze}

Following Refs.~\cite{KV2008-dynamics,KV2009-electroweak},
take quadratic and linear \emph{Ans\"{a}tze} for the
vacuum energy density and the gravitational coupling parameter:
\bsubeqs\label{eq:Ansatz-rhoV-K}
\beqa
\rho_{V}(q) &=& \frac{1}{2}\,\big( q-q_0 \big)^2\,,
\label{eq:Ansatz-rhoV}\\[2mm]
K(q) &=& \frac{1}{2}\,q \,.
\label{eq:Ansatz-K}
\eeqa
\esubeqs
These \emph{Ans\"{a}tze} imply the following equilibrium value for
the $q$--field of theory~\eqref{eq:action-Seff-q2def-Fdef}:
\beq\label{eq:q0}
q_0 = 1/(8\pi G_\text{N}) \equiv \big(E_\text{Planck}\big)^2\,,
\eeq
where $E_\text{Planck}$ is the energy scale from \eqref{eq:EPlanck-def}.
In this article, $q$ is considered to be realized by a
4-form field strength $F$ with mass dimension 2.
With a proportionality constant in \eqref{eq:Ansatz-K} of order unity,
the natural scale of the vacuum variable $q$ is then of order
$(E_\text{Planck})^2$. However, the energy scale of $q$
in the cosmological ODEs \eqref{eq:modifiedODEs} is, in principle, arbitrary,
as noted already in the second paragraph of Sec.~\ref{subsec:Modified-ODEs}.

Next, recall the time scale $t_\text{ew}$ defined
by \eqref{eq:t_ew-def}, which corresponds to the age of
the Universe at a temperature of order $E_\text{ew}$.
For later use, also define the following number characterizing
the hierarchy of energy densities:
\beq\label{eq:xi-def}
\xi  \equiv  (E_\text{Planck}/E_\text{ew})^4 \,,
\eeq
which is of order $10^{60}$ for $E_\text{ew} \sim 1\;\text{TeV}$.
For such a large value of $\xi$, the cosmic time $t_\text{ew}$
considered in this article is large compared to the Planck time,
$t_\text{ew}= \sqrt{\xi}\, (E_\text{Planck})^{-1}$.
In addition, the relation $(t_\text{ew})^2\, q_0= \xi$ can be seen to hold,
which will be used later for the derivation of the dimensionless ODEs.

With $t_\text{ew}$ and $\xi$, the following dimensionless variables
can be defined for the cosmic time, the Hubble expansion rate,
the energy densities, and the $q$ shift away from equilibrium:
\bsubeqs\label{eq:dimensionless-var}
\beqa
\tau &\equiv& (t_\text{ew})^{-1}\, t \,,
\qquad\;\;\;\;
h   \equiv  t_\text{ew}\,H \,,
\label{eq:dimensionless-var-tau+h}\\[2mm]
r_{V} &\equiv&  (t_\text{ew})^4\, \rho_{V} \,,
\qquad
r_{Mn} \equiv \xi^{-1}\, (t_\text{ew})^4\, \rho_{Mn} \,,
\label{eq:dimensionless-var-rV+rM}\\[2mm]
x &\equiv&  \xi\, \big( q/q_0 - 1\big) \equiv \xi\, y  \,,
\label{eq:dimensionless-var-x}
\eeqa
\esubeqs
where $n$ stands for the matter-species label ($n=1,\, 2$) and $y$ is the
variable used previously in Refs.~\cite{KV2008-dynamics,KV2009-electroweak}.
Observe that $\rho_{Mn}$ has been rescaled by an extra
factor $1/\xi$ but $\rho_{V}$ not.

At this moment, it is appropriate to give explicit examples for
the EOS function $\kappa_{M1}$ and  the dissipation function $\gamma$
discussed in Sec.~\ref{subsec:Basic-idea}.
With central value $\tau_{c}$ and total width $\Delta\tau$, define the
auxiliary variable $\sigma\equiv 2\,(\tau-\tau_{c})/\Delta\tau$ and
take the EOS function to be given by
\beq\label{eq:function-kappaM1}
\kappa_{M1}(\tau) =
\left\{\begin{array}{c}
  \kappa_{c}\,\sin^2\left[ (\pi/2)\, \left(1+\sigma^2 \right) \right]\;\;
  \text{for}\;\; -1\leq \sigma \leq 1 \,, \\
  0 \hspace*{29mm}  \text{otherwise}\,.
\end{array}\right.
\eeq
In the main part of this article, $\kappa_{c}$ is set to $1$ and the
dynamic strength of the kick is controlled by the value of the initial
energy-density ratio $\rho_{M1}/\rho_{M2}$
(see App.~\ref{sec:appendixA} for a more realistic EOS function).

Turning to the dissipation function $\gamma$, introduce the
basic time scale $\tau_{\infty}$,
define
\bsubeqs\label{eq:tau_freeze+function-gamma}
\beqa
\tau_\text{freeze}  &\equiv& (1+1/\zeta)\,\tau_{\infty}\,,
\label{eq:tau_freeze}
\eeqa
and take the function to be given by
\beq
\gamma(\tau)   =
\left\{\begin{array}{c}
\cos^2\left[(\pi/2) \;\tau/\tau_\text{freeze} \right]
\;\;  \text{for}\;\;0\leq  \tau \leq \tau_\text{freeze}\,,\\
  0 \hspace*{21mm}  \text{otherwise}\,.
\end{array}\right.
\label{eq:function-gamma}
\eeq
\esubeqs
The two functions used will also be shown in the plots
of the numerical results later on.

\subsection{Dimensionless ODEs}
\label{subsec:Dimensionless-ODEs}
\vspace*{0mm}

Take, now, the \emph{Ans\"{a}tze}
\eqref{eq:Ansatz-rhoV}--\eqref{eq:Ansatz-K}
and assume small deviations of $q$ away from
the equilibrium value $q_0$, i.e., $|q/q_0 - 1| \ll 1$.
Then, the ODEs \eqref{eq:modifiedODEs}
 reduce to the following four equations
for the four dimensionless variables
$h(\tau)$, $r_{M1}(\tau)$, $r_{M2}(\tau)$, and $x(\tau)$
from \eqref{eq:dimensionless-var}:
\bsubeqs\label{eq:modifiedODEs-dimensionless}
\beqa
3\,  \big(\dot{h} +2h^2 \big) &=& \gamma\,x + ( 1- \gamma)\,\xi^{-1}\, x^2 \,,
\label{eq:modifiedODEs-hdot}
\\[1mm]
\dot{r}_{M1}+(4-\kappa_{M1})\,h\,r_{M1}
&=& - (\zeta/\gamma)\;[\dot{x}]
    - \lambdaOneTwo\, ( 1- \gamma )\,r_{M1}  \,,
\label{eq:modifiedODEs-rM1dot}
\\[1mm]
\dot{r}_{M2}+4\,h\,r_{M2}&=&+\lambdaOneTwo\, ( 1- \gamma )\,r_{M1} \,,
\label{eq:modifiedODEs-rM2dot}\\[1mm]
3\,\left( \xi^{-1}\,h\,\dot{x}+ h^2 \right)&=& \xi^{-1}\,x^2/2
+r_{M1}+r_{M2} \,,
\label{eq:modifiedODEs-xdot}
\eeqa
\esubeqs
where the overdot stands for differentiation with respect to $\tau$.
Henceforth, the functions $\kappa_{M1}(\tau)$ and $\gamma(\tau)$
are considered to be given by the explicit expressions
\eqref{eq:function-kappaM1}  and \eqref{eq:tau_freeze+function-gamma}.
For the numerical calculation, the factor $[\dot{x}]$ on the right-hand side
of \eqref{eq:modifiedODEs-rM1dot} is to be replaced by the
appropriate expression for $\dot{x}$ obtained from \eqref{eq:modifiedODEs-xdot}.
Recall, furthermore, that the dimensionless vacuum energy density
$r_{V}$ is given by $x^2/2$ and that,
according to \eqref{eq:dimensionless-var-rV+rM},
the dimensionless matter energy densities
$r_{M1}$ and $r_{M2}$ include an extra numerical factor $\xi^{-1}$
compared to $r_{V}$.

For later use, also the ODEs for the special case $\xi^{-1}=0$ are
needed. From \eqref{eq:modifiedODEs-dimensionless},
the following system of equations
can be derived for $\xi^{-1}=0$, with three ODEs:
\bsubeqs\label{eq:modifiedODEs-dimensionless-xiinfty}
\beqa
3\,  \big(\dot{h} +2h^2 \big) &=& \gamma\,x \,,
\label{eq:modifiedODEs-hdot-xiinfty}
\\[1mm]
\dot{x} &=& - \zeta^{-1}\,h\,\gamma\,
\big(2\,\gamma\,x -  \kappa_{M1}\,[3\, h^2 - r_{M2}]\,\big)\,,
\label{eq:modifiedODEs-xdot-xiinfty}
\\[1mm]
\dot{r}_{M2}+4\,h\,r_{M2}&=&+\lambdaOneTwo\, ( 1- \gamma )\,[3\,  h^2-r_{M2}] \,,
\label{eq:modifiedODEs-rM2dot-xiinfty}
\eeqa
\esubeqs
and a single algebraic equation:
\beqa
3\,  h^2 &=& r_{M1}+r_{M2} \,.
\label{eq:modifiedEq-Friedmann-xiinfty}
\eeqa
The derivation of \eqref{eq:modifiedODEs-xdot-xiinfty}
proceeds in three steps: first, take the time derivative
of \eqref{eq:modifiedEq-Friedmann-xiinfty};
second, use \eqref{eq:modifiedODEs-hdot-xiinfty} to eliminate
$\dot{h}$ in the resulting expression for $(\dot{r}_{M1}+\dot{r}_{M2})$;
third, use the final expression for $(\dot{r}_{M1}+\dot{r}_{M2})$
in the sum of the two Eqs.~\eqref{eq:modifiedODEs-rM1dot}
and \eqref{eq:modifiedODEs-rM2dot}
to get \eqref{eq:modifiedODEs-xdot-xiinfty}.
All in all, the $\xi^{-1}=0$ equations consist of three ODEs
\eqref{eq:modifiedODEs-hdot-xiinfty}--\eqref{eq:modifiedODEs-rM2dot-xiinfty}
for three variables $h(\tau)$, $x(\tau)$, and $r_{M2}(\tau)$, with
the energy density   $r_{M1}$ following from the Friedmann
Eq.~\eqref{eq:modifiedEq-Friedmann-xiinfty}.
These ODEs will be used in Sec.~\ref{subsec:xi-largerthan-100} to get
the value of the remnant vacuum energy density $r_{V}$
for $\xi\to\infty$.

Purely mathematically, there is another special case to consider
for the ODEs \eqref{eq:modifiedODEs-dimensionless} as they stand,
namely, the case $\xi=0$.  From \eqref{eq:modifiedODEs-hdot}
and \eqref{eq:modifiedODEs-xdot},
together with the proper boundary condition $x(0)=0$,
the $\xi=0$ solution can be seen to have $x(\tau)=0$,
which implies $r_{V}(\tau)=0$.
But, as said, this solution is not directly relevant  for
the physical situation considered.

\subsection{Additional remarks}
\label{subsec:Additional-remarks}
\vspace*{0mm}

A few remarks may be helpful to better understand the
proposed ODEs, given by  \eqref{eq:modifiedODEs}
in the general form or by \eqref{eq:modifiedODEs-dimensionless}
in the specialized and dimensionless form.

First, note that the previous results~\cite{KV2009-electroweak}
on the dynamics at $\tau \sim 1$ are readily recovered.
 From \eqref{eq:modifiedODEs-dimensionless} for $\zeta=0$,
 $\gamma(\tau)=1$, and $\xi\, r_{M1}\gg \xi\, r_{M2}\gg r_{V}$,
one immediately obtains
at $\tau \sim 1$\bsubeqs\label{eq:classical-xkick-rVkick}
\beqa
x(\tau)&\sim& \frac{3}{2}\; \kappa_{M1}(\tau) \, h(\tau)^2\,,
\label{eq:classical-xkick} \\[2mm]
r_{V}(\tau)  &\sim& \frac{9}{8}\; \kappa_{M1}(\tau)^2 \, h(\tau)^4  \,,
\label{eq:classical-rVkick}
\eeqa
\esubeqs
which corresponds to Eqs.~(3.1a) and (3.4a) of Ref.~\cite{KV2009-electroweak}
apart from a trivial rescaling.
With  $\kappa_{M1}(\tau)^2\, h(\tau)^4$ dropping  to
zero rapidly for large enough $\tau$, there is no sizable remnant vacuum energy density,
at least, according to the unmodified ODEs given by
\eqref{eq:modifiedODEs-dimensionless} for $\zeta=0$ and $\gamma=1$.

For nonzero $\zeta$, however, the same approximations in the ODEs
\eqref{eq:modifiedODEs-dimensionless}
give the following dissipation-type equation at $\tau \sim 1$:
\bsubeqs\label{eq:xdoteq-Gamma}
\beqa
\dot{x}(\tau) &\sim& - \gamma_\text{diss}(\tau)\,
\big[\,\gamma(\tau)\,x(\tau) - (3/2)\, \kappa_{M1}(\tau) \, h(\tau)^2\,\big]\,,
\label{eq:xdoteq}\\[2mm]
\gamma_\text{diss}(\tau)  &\equiv&  2\,\zeta^{-1}\,h(\tau)\,\gamma(\tau)\,,
\label{eq:Gamma}
\eeqa
\esubeqs
whose derivation parallels the one of
\eqref{eq:modifiedODEs-xdot-xiinfty} in the previous subsection.
Equation~\eqref{eq:xdoteq} with boundary condition $x(0)=0$
is the analogue of the crucial relation (4.1) of
Ref.~\cite{KV2009-electroweak} that allows for a positive
remnant vacuum energy density as discussed in Sec.~IV of that article.
Section~IV of Ref.~\cite{KV2009-electroweak} contains, in fact,
the analytic solution of \eqref{eq:xdoteq}
for given  functions  $\gamma(\tau)$, $\kappa_{M1}(\tau)$, and $h(\tau)$.
For completeness, the dimensionful quantity corresponding
to $\gamma_\text{diss}(\tau)$ is given by
$\Gamma_\text{diss}(t)= 2\,\zeta^{-1}\,H(t)\,\gamma(t/t_\text{ew})$.

The dissipative ODE \eqref{eq:xdoteq-Gamma}
and its analytic solution~\cite{KV2009-electroweak}
make clear that a finite remnant vacuum energy density with
$\lim_{\tau\to\infty} x(\tau) \ne 0$
requires both $\zeta> 0$ and $\zeta <\infty$. Indeed,
for the case $\gamma_\text{diss}\to\infty$ (or $\zeta\to 0$),
the solution $x(\tau)$ follows $\kappa_{M1}(\tau)\,h(\tau)^2$
which drops to zero rapidly for large times and,
for the case $\gamma_\text{diss}\to 0$ (or $\zeta\to \infty$),
the solution  $x(\tau)$ simply remains at the initial
value, $x(\tau)=x(0)=0$.

Second, it appears essential that the
ODEs \eqref{eq:modifiedODEs} and \eqref{eq:modifiedODEs-dimensionless}
are \emph{singular}, with the coefficients of the first terms on the
right-hand sides
of \eqref{eq:modifiedODEs-M1dot} and \eqref{eq:modifiedODEs-rM1dot}
diverging for $\tau > \tau_\text{freeze}$,
because of the $\gamma$ condition \eqref{eq:gamma-conditions-bc-larget}.
In fact, the divergence of the coefficient $\zeta/\gamma(\tau)$ for
$\tau$ above $\tau_\text{freeze}$ forces $(q/q_0-1)$
to be \emph{strictly} constant.
It could very well be that the exact vanishing of $\gamma$
for $\tau > \tau_\text{freeze}$ in the
cosmological context traces back to the existence of an energy threshold
in the relevant particle reaction process.
As noted in Ref.~\cite{KV2009-electroweak}, the energies involved
are tiny (of the order of meV), so that only sufficiently light neutrinos
and gravitons can be expected to play a role. The first term
on the right-hand side of \eqref{eq:modifiedODEs-M1dot}, as it stands,
is not simply proportional to $R^2$ as for the well-known
Zeldovich--Starobinsky result~\cite{ZeldovichStarobinsky1977},
but does involve $R$ via its time derivative,
as follows by use of \eqref{eq:modifiedODEs-Hdot}.

Third, the discussion of the two previous remarks suggests that
the value of the remnant vacuum energy density can be at most
of the order of the maximum possible ``classical'' value (i.e.,
the $r_{V}$ peak from the nondissipative $\zeta=0$ equations).
The idea is that, in general, dissipation leads to
reduction of the produced energy rather than enhancement.
Specifically, the conjectured inequality is
\beq\label{eq:rV-upperbound}
r_{V}(\tau_\text{freeze}) \lesssim
\max_{\tau}\,\left[\frac{9}{8}\;\kappa_{M1}(\tau)^2 \,
\big(h(\tau)^2 - r_{M2}(\tau)/3\big)^2 \right]\,,
\eeq
which is based on the analytic result \eqref{eq:classical-rVkick} with
$h^4$ on the right-hand side replaced by $(h^2 - r_{M2}/3)^2$,
as suggested by \eqref{eq:modifiedODEs-xdot-xiinfty}.
It remains to sharpen the approximate upper bound \eqref{eq:rV-upperbound}
and to determine the corresponding conditions.

Fourth, having a constant nonzero value of $x \propto (q/q_0-1)$
does not automatically allow for a standard de-Sitter universe,
as the original ODE \eqref{eq:modifiedODEs-hdot} [with $\gamma\equiv 1$]
and the ODE \eqref{eq:modifiedODEs-xdot}
are inconsistent for $\dot{x}=\dot{h}=0$, $x < \xi$,  and $r_{M1}=r_{M2}=0$.
However, the modified Eq.~\eqref{eq:modifiedODEs-hdot}
[with $\gamma(\tau)=0$ for $\tau > \tau_\text{freeze}$]
has been designed to
match the corresponding Einstein equation of a standard flat FRW model
with ultrarelativistic matter and constant vacuum energy density,
which asymptotically approaches a de-Sitter universe.\footnote{The assumption,
here, is that other contributions to the vacuum energy density generated at times
later than $t_\text{ew}$ would be self-adjusted away by appropriate
$q$--type fields~\cite{KV2008-statics}.
The prime example is the quantum-chromodynamics (QCD)
vacuum energy density of order $(100\;\text{MeV})^4$, which is
expected to appear during the cosmological QCD transition
at $T\sim 100\;\text{MeV}$. This huge contribution to the
vacuum energy density $\rho_{V}$ has been shown to self-adjust
to zero~\cite{KV2009-gluoncondcosm,Klinkhamer2009},
as long as there is no term proportional to $|H| E_\text{QCD}^3$
contributing to $\rho_{V}$.
If there is such a nonanalytic term (cf. the discussion
in Ref.~\cite{UrbanZhitnitsky2009}),
then the final value of $\Lambda$ could be a combination of
electroweak and QCD effects. However,
the experimental signatures of the electroweak vacuum energy density
(effectively, a $\Lambda$CDM model,
as explained in the next footnote)
and the QCD vacuum energy density
(an $f(R)$ modified-gravity model)
are different, in particular as regards the
effective EOS parameter $\overline{w}_{X}$
discussed in Ref.~\cite{Klinkhamer2009}.
But, for the moment, there is no definitive proof that the
required nonanalytic term occurs in four-dimensional QCD.\label{ftn:other-kicks}}
This particular modification
also makes clear that there must be more than just
energy exchange between the vacuum and matter sectors.
Rather, there must be a type of \emph{modulated interaction}
between  the vacuum field and the nonstandard
gravitational field, which can be seen as follows.
Multiply \eqref{eq:modifiedODEs-Hdot}
by $K\,dq/dK$ to get a modified FRW--Einstein equation,
\vspace*{1mm}
\beqa
6\,K \left(\frac{d H}{d t} +2H^2 \right)
&=&
\big[\gamma(t/t_\text{ew})\big]\,K\,\frac{d\rho_{V}}{d K}
+ \big[ 1- \gamma(t/t_\text{ew})  \big]\,2\rho_{V}\,,
\eeqa
where the nonstandard term $K\,d\rho_{V}/d K$
is switched off for large enough cosmic times
by the factor $\gamma(t/t_\text{ew})$ going to zero.

Fifth, the ODEs \eqref{eq:modifiedODEs-dimensionless},
for given functions $\kappa_{M1}(\tau)$ and $\gamma(\tau)$
and fixed coupling constants $\zeta$ and $\lambdaOneTwo$,
contain one last free parameter, the hierarchy parameter $\xi$
defined by \eqref{eq:xi-def}.
Heuristically, it is to be expected that the precise value of $\xi\gg 1$
does not affect the resulting
value $r_{V}(\tau_\text{freeze})$. But, as $\xi$ fixes the ratio
$\rho_\text{M}/\rho_{V}$ at $t\sim t_\text{ew}$,
it does affect the later (standard) evolution of the model universe
and, in fact, determines~\cite{ArkaniHamed-etal2000}
the cosmic time $t_\text{accel}$ at which the matter energy density
drops below that of the constant vacuum energy density,
$t_\text{accel} \sim \sqrt{\xi}\,t_\text{ew}$.
More precisely, the onset of acceleration $\ddot{a}/a > 0$
[having defined $h\equiv \dot{a}/a$ in terms of the scale factor $a(\tau)$]
occurs at the energy-density ratio $\rho_{V}/\rho_\text{Mtot} = \overline{\upsilon}$,
with $\overline{\upsilon}=1$
for relativistic matter ($w_{M2}=1/3$ and $\kappa_{M2}=0$)
or $\overline{\upsilon}=1/2$ for
nonrelativistic matter ($w_{M2}=0$ and $\kappa_{M2}=1$).
The model considered in the present article has $\overline{\upsilon}$ $=$ $1$,
but can easily be adapted to give the value $\overline{\upsilon}$ $=$ $1/2$
which is more realistic.\footnote{The
adapted model contains an additional EOS function
$\kappa_{M2}(t/t_\text{eq})$, which is a smoothed
step function running from $0$ to $1$ as the cosmic time $t$ increases
and which has a half-way time $t_\text{eq}$ where
$\kappa_{M2}(1)=1/2$.
This matter-radiation-equality time $t_\text{eq}$ has a
parametric form  $\alpha^3\,\xi^{1/2}\,t_\text{ew}$, with
$\alpha$ the electromagnetic fine-structure constant
(see Ref.~\cite{ArkaniHamed-etal2000} for further details).
The adapted model then has an acceleration phase for
$\rho_{V}/(\rho_{M1}+\rho_{M2}) \approx \rho_{V}/\rho_{M2} > 1/2$
and corresponds, for $t\gg t_\text{ew}$,
to a particular $\Lambda$CDM model~\cite{Komatsu2008,Mukhanov2005}.
As noted in Ref.~\cite{KV2009-electroweak},
this EOS function $\kappa_{M2}$ can also be expected to perturb the
vacuum energy density,
but the magnitude involved is tiny compared to the one from the electroweak scale
because $H(t_\text{eq})^4 \ll H(t_\text{ew})^4$.\label{ftn:adapted-model}}

\section{Numerical solution}
\label{sec:Numerical-solution}

\subsection{Numerical results for $\boldsymbol{\xi=10^2}$}
\label{subsec:xi100}

\vspace*{-2mm}
The mathematical parameter $\xi$ entering the
ODEs \eqref{eq:modifiedODEs-dimensionless}
is, first, considered to have the moderately
large value of 100.  The boundary conditions are taken from the epoch
before the electroweak kick,
when there was a standard radiation-dominated flat FRW universe.
With the onset of the electroweak kick given by $\tau_\text{start}$
from \eqref{eq:kappaM1-conditions-nonzero} and \eqref{eq:function-kappaM1},
the following boundary conditions
on $h$, $x$, and $r_\text{Mtot}\equiv r_{M1} + r_{M2}$
hold at a time $\tau=\tau_\text{min} \leq \tau_\text{start}<\tau_\text{ew}$:
\bsubeqs\label{eq:modifiedODEs-dimensionless-bcs}
\beqa
h(\tau_\text{min})&=& 1/2 \;(\tau_\text{min})^{-1} \,,\\[2mm]
x(\tau_\text{min})&=& 0 \,,\\[2mm]
r_\text{Mtot}(\tau_\text{min})&=& 3 \;\big[h(\tau_\text{min})\big]^2\,.
\eeqa
\esubeqs
This leaves only the initial ratio
$[r_{M1}/r_{M2}](\tau_\text{min})\equiv
r_{M1}(\tau_\text{min})/r_{M2}(\tau_\text{min})$ to be determined,
which is, for the moment, taken to be $1$
(other initial ratios will be discussed shortly).

\begin{figure*}[t]           
\vspace*{0mm}
\includegraphics[width=\textwidthPreprintTwocolumn]{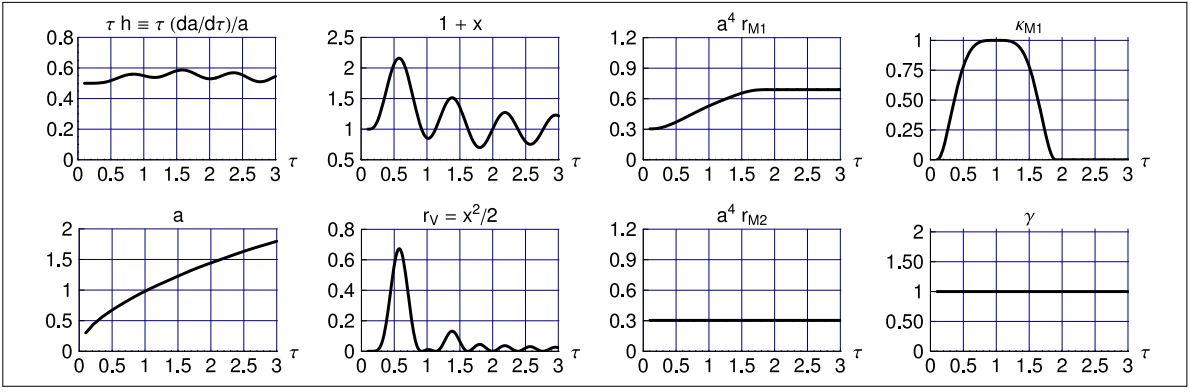}
\vspace*{3mm}
\caption{Numerical solution of the ODEs \eqref{eq:modifiedODEs-dimensionless}
with a prescribed EOS function $\kappa_{M1}$,
for vanishing  dissipative coupling constant $\zeta=0$
and trivial function $\gamma(\tau)=1$.
Model parameters are $\{\xi,\,   \lambdaOneTwo \}$ $=$
$\{100,\,   8/100\}$
and the model function $\kappa_{M1}(\tau)$ is
defined by \eqref{eq:function-kappaM1} with parameters
$\{\kappa_{c},\, \tau_{c},\, \Delta\tau \}$ $=$ $\{1,\,1,\,18/10\}$.
The ODEs are solved over the interval
$[\tau_\text{min},\, \tau_\text{max}]$ $=$ $[0.1,\,  3]$ with the following
boundary conditions from \eqref{eq:modifiedODEs-dimensionless-bcs}
at $\tau=\tau_\text{min}=0.1$:
$\{x,\, h,\, a,\, r_{M1},\, r_{M2}\}$ $=$
$\{0,\,  5,\,  0.3,\,  37.5,\, 37.5 \}$.}
\label{fig:1} \vspace{0cm}
\end{figure*}

The corresponding numerical
solutions of the ODEs \eqref{eq:modifiedODEs-dimensionless}
are shown in Figs.~\ref{fig:1}--\ref{fig:3}.
\mbox{Figure~\ref{fig:1}} illustrates the fact that
the standard (nondissipative) dynamic equations for $\zeta=0$
and $\gamma=1$ do not produce a constant positive remnant
vacuum energy density from the electroweak kick
[the oscillatory effects in $r_{V}(\tau)$
are suppressed for larger values of the hierarchy parameter $\xi$;
see the first figure called in Sec.~\ref{subsec:xi-largerthan-100}].
However, as shown by Fig.~\ref{fig:2},
the modified (dissipative) dynamic equations for $\zeta=\text{O}(1)$
do produce a sizable remnant vacuum energy density.

The subsequent evolution of Fig.~\ref{fig:2} is shown in
Fig.~\ref{fig:3}. The content of this model universe for $\tau >  3$
is given by a constant vacuum energy density
(effective cosmological constant)
and two species of matter, with massive type--1 particles playing a role for the
generation of the vacuum energy density during the electroweak epoch
($\tau \sim 1$) and ultimately decaying into massless type--2 particles.

Similar results are obtained for initial ratios
$[r_{M1}/r_{M2}](\tau_\text{min}) \gtrsim 1$.
Table~\ref{tab-rM1rM2-xi100} gives the
function values at cosmic time $\tau=\tau_\text{freeze}$,
in particular, the values for $r_{V}$ which, by construction,
stay constant for later times $\tau > \tau_\text{freeze}$.
Remark that even for a relatively mild kick with initial ratio
$[r_{M1}/r_{M2}] = 1/10$, the generated $r_{V}$ is still
of order $10^{-3}$.

Returning to the boundary condition $[r_{M1}/r_{M2}](\tau_\text{min}) = 1$,
Figs.~\ref{fig:2}--\ref{fig:3} are seen to give a value
$r_{V}(\tau_\text{freeze})\approx 0.04$.
By changing the model parameters and the model functions somewhat it
is possible to get $r_{V}(\tau_\text{freeze})$
values in the range of $10^{-3}$ to $1$.
But it appears impossible to get a remnant $r_{V}$ much larger than
unity, which agrees with the conjectured upper bound \eqref{eq:rV-upperbound}.

\subsection{Numerical results for $\boldsymbol{\xi\gg 10^2}$}
\label{subsec:xi-largerthan-100}
\vspace*{0mm}

The parameter $\xi$ has been defined in physical terms
by \eqref{eq:xi-def} and its mathematical role for the
solution of the ODEs \eqref{eq:modifiedODEs-dimensionless}
has already been discussed in the last paragraph of
Sec.~\ref{subsec:Additional-remarks}.
Here, numerical results are presented for large values
of this parameter, ranging from $\xi = 10^{4}$ to
$\xi = \infty$.

\begin{figure*}[t]
\vspace*{-.0cm}
\includegraphics[width=\textwidthPreprintTwocolumn]{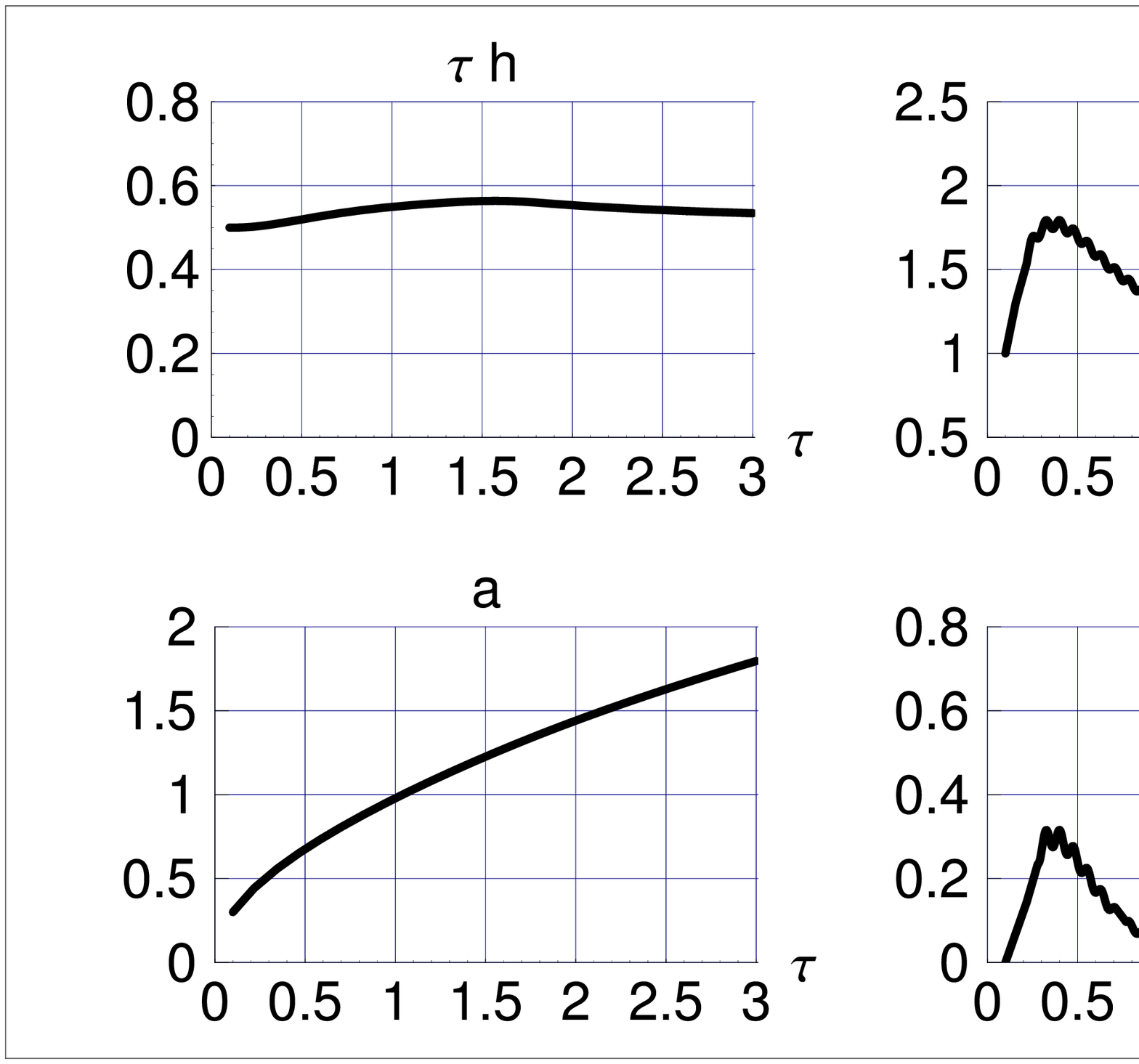}
\vspace*{3mm}
\caption{Same as Fig.~\ref{fig:1}, still with $\zeta=0$
but now for a larger hierarchy parameter $\xi=10^{4}$.}
\label{fig:4} \vspace{0mm}
\end{figure*}

Numerical results for the standard nondissipative ($\zeta=0$)
dynamic equations at $\xi = 10^{4}$
are given in Fig.~\ref{fig:4}, which show reduced
oscillatory effects of $r_{V}(\tau)$ compared to
Fig.~\ref{fig:1} and recover the smooth behavior of
\eqref{eq:classical-rVkick}. [Recall that the analytic approximation
\eqref{eq:classical-rVkick} was derived for a negligible type--2
matter energy density
and a better approximation has $h^4$ on the right-hand side replaced
by $(h^2 - r_{M2}/3)^2$, as used already in \eqref{eq:rV-upperbound}.]
Further numerical results for $\zeta=2$ and $\xi = 10^{4}$
confirm the expectation from Sec.~\ref{subsec:Additional-remarks}
that the generation of the
remnant vacuum energy density at $\tau\sim 1$ is qualitatively
the same as for $\xi = 10^{2}$
(compare Fig.~\ref{fig:5} with Fig.~\ref{fig:2})
and that the main effect of a larger value of $\xi$
is that of pushing the onset of the vacuum-dominated expansion to larger
values of $\tau$, with $\tau_\text{accel} \propto \sqrt{\xi}$
(compare Fig.~\ref{fig:6} with Fig.~\ref{fig:3}).
It is also instructive to contrast the behavior of
the vacuum variable $x(\tau)$ in Fig.~\ref{fig:4} and that
of Fig.~\ref{fig:5}, where the latter figure
displays the ``time-lag effect'' because of
the finite dissipative coupling constant $\zeta$
(cf. the heuristic discussion of the paragraph starting a few
lines under Eq.~(4.5) in Ref.~\cite{KV2009-electroweak}).

\begin{figure*}[t]   
\vspace*{0cm}
\includegraphics[width=0.275\textwidth]{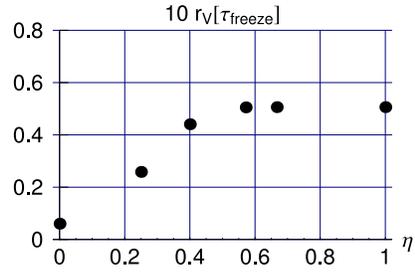}
\vspace*{3mm}
\caption{Remnant vacuum energy density $r_{V}(\tau_\text{freeze})$
from Table~\ref{tab-xi-extrapol} plotted against the compact
parameter $\eta \equiv \log_{10}\xi/(|\log_{10}\xi| +3) \in [0,\, 1)$
defined in terms of the hierarchy parameter $\xi \in [1,\, \infty)$
of the theory. The realistic value $\xi = 10^{60}$ from \eqref{eq:xi-def}
corresponds to $\eta= 20/21 \approx 0.9524$.}
\label{fig:7}\vspace*{20cm}
\end{figure*}

As the model evolution for $\tau>\tau_\text{freeze}=3$ is
perfectly standard (described by an FRW universe with
ponderable matter and an effective cosmological constant),
the crucial segment of the numerical solution is over
the interval $[\tau_\text{min},\, \tau_\text{freeze}]$.
The numerical data for the function values are given in
Table~\ref{tab-xi-extrapol}, where the $\xi=\infty$ values
refer to the solution of the equations
\eqref{eq:modifiedODEs-dimensionless-xiinfty}
derived in Sec.~\ref{subsec:Dimensionless-ODEs}.
The functions from Table~\ref{tab-xi-extrapol} are observed to converge
for $\xi \gtrsim 10^{4}$. The convergence of the
vacuum energy density results is also shown by Fig.~\ref{fig:7}.
With $r_{V}(\tau_\text{freeze})$ values for parameters $\xi$ both
below and above the ``realistic'' number $\xi = 10^{60}$,
the following estimate is obtained by interpolation:
\beq\label{eq:rV-final}
r_{V}(\tau_\text{freeze})\,
\Big|_\text{Table~\ref{tab-xi-extrapol}}^{\xi=10^{60}} \approx 0.051\,,
\eeq
for the model parameters and boundary conditions
mentioned in the caption of Table~\ref{tab-xi-extrapol}.

The function values at $\tau_\text{freeze}$ calculated from the
$\xi^{-1}=0$ ODEs \eqref{eq:modifiedODEs-dimensionless-xiinfty}
are given in Table~\ref{tab-rM1rM2-xiinfty} for a wide range of values
of the initial ratio $[r_{M1}/r_{M2}]$ at $\tau=\tau_\text{min}=0.1$.
These function values can be expected to approximate
the physical ($\xi^{-1} \sim 10^{-60}$) values with an accuracy
of one per mill or better, at least, for initial ratios $[r_{M1}/r_{M2}]$
of order unity.

As far as the dimensionless remnant vacuum energy density
$r_{V}(\tau_\text{freeze})$ is concerned, the values quoted in Table~\ref{tab-rM1rM2-xiinfty}
constitute the \emph{complete} solution of the problem where the
initial ratio $[r_{M1}/r_{M2}](\tau_\text{min})$ controls the
relative strength of the kick at $t\sim t_\text{ew}$, assuming
the validity of the phenomenological ODEs \eqref{eq:modifiedODEs}
for coupling constant $\zeta=\text{O}(1)$ and taking
model functions $\kappa_{M1}(\tau)$ and $\gamma(\tau)$
from \eqref{eq:function-kappaM1}  and \eqref{eq:tau_freeze+function-gamma},
respectively.
The dimensionful remnant vacuum energy density $\rho_{V}(t_\text{freeze})$
requires knowledge of the absolute energy scale $E_\text{ew}$ used in the rescaling
of the variables, as will be discussed further in the next section.

\section{Discussion}
\label{sec:Discussion}

\subsection{General case}
\label{subsec:General-case}

In the scenario considered~\cite{ArkaniHamed-etal2000,KV2009-electroweak},
the theoretical value of the effective cosmological constant
(remnant vacuum energy density) is given by
\beq\label{eq:Lambda-theory}
\Lambda^\text{theory} \equiv \lim_{t\to\infty}\;
\rho_V^\text{theory}(t) =r_V^\text{num}
\;(E_\text{ew})^{8}\, /(E_\text{Planck})^{4} \,,
\eeq
with $r_V^\text{num}$ a number obtained by numerically solving ODEs of the
type of \eqref{eq:modifiedODEs}. Equating
the theoretical value \eqref{eq:Lambda-theory} with the experimental
value $\Lambda^\text{exp}\approx (2\;\text{meV})^4$ from observational
cosmology~\cite{Riess-etal1998,Perlmutter-etal1998,Komatsu2008},
the following estimate for the required energy scale is obtained:
\beqa\label{eq:Eew-estimate}
E_\text{ew} &=& (\Lambda^\text{exp}/r_V^\text{num})^{1/8}\, (E_\text{Planck})^{1/2}
\approx  
3.2\;\text{TeV}\;
\left(\, \frac{0.051}{r_V^\text{num}}\, \right)^{1/8}
\left(\, \frac{\Lambda^\text{exp}}{(2.0\;\text{meV})^4} \,\right)^{1/8} \,.
\eeqa
For the moment, the numerical value $r_V^\text{num}=0.051$
in \eqref{eq:Eew-estimate} is purely for illustrative purpose.

Clearly, the calculation of the present paper relies on many
assumptions, but it appears that values of the order of unity
for the dimensionless energy density $r_V^\text{num}$ are quite reasonable.
The value $r_V^\text{num}=1.0$ corresponds to $E_\text{ew} \approx 2.2\;\text{TeV}$,
according to \eqref{eq:Eew-estimate}.
On the other hand,  values $r_V^\text{num}\gg 1$ appear unlikely,
at least, in the present framework [see
the discussion in Sec.~\ref{subsec:Additional-remarks}
leading up to \eqref{eq:rV-upperbound}].
Note that an $r_V^\text{num}$ value
of order $10^{10}$ would be required in \eqref{eq:Eew-estimate}
to bring $E_\text{ew}$ down to the order of magnitude of
standard-model particle masses, $m_\text{SM} = 10^2\;\text{GeV}$.

Taking $r_V^\text{num}\lesssim 1$ for granted,
the correct reading of \eqref{eq:Eew-estimate} is that of a lower bound,
\beq\label{eq:Eew-lower-bound}
E_\text{ew} \gtrsim 2\;\text{TeV}\,,
\eeq
since, without further input,
the value of $r_V^\text{num}$ can be made arbitrarily small
(for example, by taking a sufficiently small value for the
initial energy density $r_{M1}$ in Table~\ref{tab-rM1rM2-xiinfty}).
Indeed, the main uncertainty (apart from the unknown $E_\text{ew}$ value)
is the dynamic importance at $t\sim t_\text{ew}$ of the
nonrelativistic ($M \sim E_\text{ew}$) type--1 particles compared
to that of the relativistic  type--2 particles.

\subsection{Special case}
\label{subsec:Special-case}

In order to get further  predictions, the following three assumptions
can be made.
First, assume the type--1 and type--2 particles to have been
ultrarelativistic and in thermal equilibrium for $T\gg E_\text{ew}$,
so that their energy densities are given by
\bsubeqs
\beq
\rho_{Mn}= (\pi/30)\,N_{\text{eff},n}\,T^4\,,
\eeq
with bosons ($b$) and fermions ($f$) of particle type $n=1,2$
contributing as follows:
\beq
N_{\text{eff},n}= \sum_{b}\,g_{n,b} + (7/8)\,\sum_{f}\,g_{n,f}\,,
\eeq
in terms of the numbers of degrees of freedom
of the particles ($g_{b}=2$ for the photon).
Then, the relevant energy-density ratio $\rho_{M1}/\rho_{M2}$
before the kick starts is simply given by the ratio of the
respective effective numbers of degrees of freedom,
$N_{\text{eff},1}/N_{\text{eff},2}$.

Second, assume the type--1 particles to have approximately
the same mass and a mass scale
\beq
M \sim E_\text{ew}\,.
\eeq
As discussed in Sec.~\ref{subsec:Basic-idea}, these
type--1 particles can be thought to consist of a mix of
different unstable particles. What matters here, though,
is their average thermodynamic properties as given
by the prescribed EOS function $\kappa_{M1}$
from \eqref{eq:function-kappaM1}.
See App.~\ref{sec:appendixA} for a realistic setup.

Third, assume the massless type--2 particles to correspond to those of
the standard model ($m \lesssim m_\text{SM} \ll E_\text{ew}$), so that
\beq
N_{\text{eff},2}= N_{\text{eff,\,SM}}= 427/4 \sim 10^2\,.
\eeq
\esubeqs
See, e.g., Ref.~\cite{Amsler-etal2008} for the count
of the degrees of freedom in the standard model.

With these assumptions, there are only two unknowns:
the numerical value of the energy scale $E_\text{ew}$ and the
effective number $N_{\text{eff},1}$ of type--1 degrees of freedom.
The first quantity, $E_\text{ew}$, sets,
according to \eqref{eq:t_ew-def}, the physical time $t\sim t_\text{ew}$
when the `kick' of the vacuum energy density
occurs (the kick mechanism~\cite{KV2009-electroweak} relies
on the change of the Hubble expansion rate by type--1 mass effects).
The second quantity, $N_{\text{eff},1}$, controls the
initial energy-density ratio:
\beq
[r_{M1}/r_{M2}](\tau_\text{min})
= N_{\text{eff},1}/N_{\text{eff},2}
\sim  N_{\text{eff},1}/10^2\,,
\eeq
where the dimensionless cosmic time $\tau_\text{min}$
is taken before the kick starts
and $r$ denotes the dimensionless energy density
according to \eqref{eq:dimensionless-var}.

For a substantial number $N_{\text{eff},1}=10^2$ of these ultramassive
type--1 particles (possibly corresponding to partners of the standard-model
particles from broken supersymmetry~\cite{WessZumino1974,FayetFerrara1976}),
the initial energy-density ratio is given by
$[r_{M1}/r_{M2}](\tau_\text{min}) \sim 1$
and the dimensionless remnant vacuum energy density $r_V^\text{num}$ is
found to be of order $5.1 \times 10^{-2}$ [see Table~\ref{tab-rM1rM2-xiinfty}].
This particular $r_V^\text{num}$ value requires,
according to \eqref{eq:Eew-estimate},
an $E_\text{ew}$ value of order $3.2\;\text{TeV}$, in order
to reproduce the experimental value of the cosmological constant.
A similar $E_\text{ew}$ value is obtained in App.~\ref{sec:appendixA}
if the prescribed (artificial)  EOS function \eqref{eq:function-kappaM1}
is replaced by a physically-motivated EOS function.
Only for $N_{\text{eff},1}=1$ (corresponding to a single ultramassive
real scalar)
does the remnant vacuum energy density drop to such a low value,
$r_V^\text{num} \sim 2.0 \times 10^{-5}$ [Table~\ref{tab-rM1rM2-xiinfty}],
that the required energy scale becomes significantly larger,
$E_\text{ew} \sim 8.5\;\text{TeV}$.

Hence, if all particles have initially been in thermal equilibrium
and if the type--1 particles with mass scale $M \sim E_\text{ew}$
have an effective number of degrees of
freedom $N_{\text{eff},1} \gtrsim 1$,
the required energy scale $E_\text{ew}$
from \eqref{eq:Eew-estimate} lies in the following range:
\beq\label{eq:Eew-estimate-prescribed-kick}
E_\text{ew}\;
\big|^\text{\,prescribed kick}_{N_{\text{eff},1}\geq 1,\;N_{\text{eff},2}=10^2} \sim 3 - 9\;\text{TeV}\,,
\eeq
assuming a dissipative coupling constant $\zeta$ of order unity.
The trend is, as expected from \eqref{eq:Lambda-theory}, that a smaller
number $N_{\text{eff},1}$ requires a larger energy scale $E_\text{ew}$.
Table~\ref{tab-Neff1-Eew} presents the required energy scales $E_\text{ew}$
for selected values of $N_{\text{eff},1}$, with the understanding
that the quoted numbers for $E_\text{ew}$ are only indicative
because of the many assumptions made along the way
(some of which may be more reasonable than others).
See App.~\ref{sec:appendixA} for further discussion of some of
the systematic uncertainties involved (its very last table complements
Table~\ref{tab-Neff1-Eew} of this section).

\subsection{Outlook}
\label{subsec:Outlook}

If the observed  ``cosmological constant'' results from dynamics at
cosmic temperatures of order $M \sim E_\text{ew}$, then
\emph{some} set of differential equations must be relevant.
In the framework of $q$--theory, a particular
set of differential equations has been proposed.
It appears to be impossible to have very much simpler differential
equations which achieve the same result and
the phenomenological equations used here
can be expected to carry some of the essential ingredients.
If so, the estimates from Table~\ref{tab-Neff1-Eew} suggest
the need for new physics with particle masses at the TeV--scale.

Particle-collider experiments are called upon to confirm or exclude
the existence of these TeV--scale particles and, if confirmed,
to determine their characteristics.
Knowing the characteristics of the new TeV-scale particles
(assuming their detection), the main task for theorists
would be to \emph{derive} the relevant particle-production effects
contained in the simple phenomenological equations considered here
or to find the appropriate generalizations of these equations.

\section*{\hspace*{-4.5mm}ACKNOWLEDGMENTS}
\noindent It is a pleasure to thank G.E. Volovik for numerous
discussions on vacuum energy over the years and the referee
for helpful comments on earlier versions of this article.

\begin{appendix}
\section{Dynamic kick}
\label{sec:appendixA}

\subsection{Introduction}
\label{subsec:appA-Introduction}

The goal of this appendix is to present a
calculation for the generation of the remnant vacuum energy density
by a more or less realistic kick from dynamically generated
ultramassive and unstable type--1 particles.
The description of this dynamic kick is rather involved, but the
final ODEs are similar to those of the main text for a prescribed
kick, with the crucial role again being played by the phenomenological
dissipation function $\gamma$
(the other functions entering the ODEs will now be given
different notations, for clarity).
The heuristics of these ODEs will be discussed in the
last paragraph of Sec.~\ref{subsec:appA-ODEs-and-BCS}.

\subsection{Mass spectrum and EOS functions}
\label{subsec:appA-Mass-spectrum-and-EOS}

The massless type--2 particles from the model introduced
in Sec.~\ref{subsec:Basic-idea} are considered to correspond to those of
the standard model (mass scale $m_\text{SM} \ll E_\text{ew}$)
and the round number
$N_2\equiv N_{\text{eff},2} =10^2$ will be used.
The ultramassive type--1 particles are considered to arise from
broken supersymmetry~\cite{WessZumino1974,FayetFerrara1976},
so that $N_1\equiv N_{\text{eff},1} =N_{\text{eff},2}=10^2$,
and they can be taken to be bosons (the standard-model
particles being mostly fermions~\cite{Amsler-etal2008}).
Just as discussed in Sec.~\ref{subsec:Special-case},
the type--1 and type--2 particles
are assumed to have been in thermal equilibrium
before the generation of the vacuum energy density starts.

A general type--1 mass spectrum is given by the effective
numbers $n_{i}\geq 0$ of particles with dimensionless
masses $m_{i}\equiv M_i/E_{ew}$, for which the following
two constraints hold:
\bsubeqs\label{eq:general-mass-spectrum-AppNew}
\beqa
\sum_{i=a,b,c,\ldots}\; n_{i}                    &=& N_{1}\,,\\[2mm]
\frac{1}{N_{1}}\;\sum_{i=a,b,c,\ldots}\; n_{i}\, m_{i}&=& 1\,,
\eeqa
\esubeqs
where the last constraint ensures that the average dimensionful mass
equals $E_{ew}$.

For simplicity, consider two cases: case A with two different mass
values and case B with a single mass value, by definition equal to $E_{ew}$.
(The generalization to a general type--1 mass spectrum will be obvious.)
The specific numbers are chosen as follows:%
\bsubeqs\label{eq:caseAandB-mass-spectrum-AppNew}
\beqa
\text{case\;A:}\quad  (n_{a},\,m_{a}\,;\,n_{b},\,m_{b}) &=&
(\phantom{0}40,\,2\,;\,60,\,1/3)\,,
\label{eq:caseA-mass-spectrum-AppNew}\\[2mm]
\text{case\;B:}\quad  (n_{a},\,m_{a}\,;\,n_{b},\,m_{b}) &=&
\label{eq:caseB-mass-spectrum-AppNew}
\eeqa
\esubeqs
The case--B partition $\{n_{a},\,n_{b}\}$ of 100 is arbitrary,
at least, for the dynamic ODEs considered here.
It will be seen, later on, that case A and case B give more or less
the same remnant vacuum energy density, which even holds for the
extreme version of case--A having 50 particles with $m_{a}=2$
and 50 with $m_{b}=0$.

Next, take, instead of the prescribed
EOS function \eqref{eq:function-kappaM1}
used in the main text, the  \emph{exact} EOS function or, at least,
a controlled approximation of it. In fact, the following rational
function of the variable $\theta_i\equiv T/M_i$ will be employed
for type--1 subspecies $i\in \{a,\,b,\,c,\,\ldots \}$:
\bsubeqs\label{eq:kappaM-wM-alphabeta-theta-AppNew}
\beqa
\overline{\kappa}_{M1i}(\theta_i) &=& 1- 3\:\overline{w}_{M1i}(\theta_i)\,,
\label{eq:kappaM-AppNew}\\[2mm]
\overline{w}_{M1i}(\theta_i) &=&
\frac{\theta_i^2+\overline{\alpha}\,\theta_i}
     {3\,\theta_i^2+\overline{\beta}\,\theta_i+\overline{\alpha}}\,,
\label{eq:wM-AppNew}\\[2mm]
\big(\overline{\alpha},\, \overline{\beta}\,\big) &=& \big(0.625,\, 1.91\big)\,.
\label{eq:alphabeta-AppNew}
\eeqa
Expression \eqref{eq:wM-AppNew} with constants \eqref{eq:alphabeta-AppNew}
gives an accurate approximation  (better than 1 per mill)
to the exact equilibrium result for the EOS parameter
$w_{M1i}\equiv P_{M1i}/\rho_{M1i}$ of bosons at zero chemical potential
(see, e.g., Sec.~3.3 of Ref.~\cite{Mukhanov2005}).
For the EOS parameter $w_{M1i}$ of fermions, the constants would be
$(\overline{\alpha},\, \overline{\beta}\,) = (0.770,\, 2.15)$, but the
approximation would be somewhat less accurate
(still better than 5 per mill).

As the type--2 particles are massless and adiabatically expanding in the
early phase (for $\tau \leq \tau_\text{min}\ll 1$), their initial energy density
can be used to normalize the temperature-over-mass variable $\theta_i$
used in \eqref{eq:kappaM-AppNew},
\beqa\label{eq:theta-AppNew}
\hspace*{-9mm}
\theta_i &=& \frac{1}{m_{i}\,E_\text{ew}}
\left(\frac{30}{\pi\,N_{\text{eff},2}}\;\rho_{M2}(t_\text{min})\right)^{1/4}
\frac{a(t_\text{min})}{a(t)}
=   
\frac{1}{m_{i}}
\left(\frac{30}{\pi\,N_{\text{eff},2}}\;r_{M2}(\tau_\text{min})\right)^{1/4}
\frac{a(\tau_\text{min})}{a(\tau)}\,,
\eeqa
\esubeqs
where $m_{i}\,E_\text{ew}$ is the physical mass $M_i$ of the type--1 subspecies
considered. As the scale factor $a(\tau)$
evolves with dimensionless cosmic time $\tau$,
so do $\theta_i$ and $\overline{\kappa}_{M1i}$ in the above equations.
Recall that the scale factor $a(\tau)$ is defined by $h=\dot{a}/a$
and that the temperature of a noninteracting gas of ``photons''
drops as $T(\tau) \propto 1/a(\tau)$ in the expanding FRW
universe~\cite{Mukhanov2005}.

\subsection{Model functions}
\label{subsec:appA-Model-functions}

Because the EOS function $\overline{\kappa}_{M1i}$ does not vanish
for $\tau>\tau_\text{freeze}$,
the appropriately modified Maxwell equation needs to be used, which
has already been discussed in Ftn.~\ref{ftn:changed-mod-Maxwell-eq}.
But $\overline{\kappa}_{M1i}$ also does not vanish exactly
in the early phase (as long as $a>0$ or $T<\infty$).
In order to have a standard radiation-dominated FRW universe in the
early phase before $t_\text{ew}$, the initial density of type--1
particles must be taken to be strictly zero.
The further assumption is, then, that type--1 particles are generated
dynamically just before $t_\text{ew}$
and thermalized rapidly, so that the type--$1i$ energy density
relative to that of the type--2 particles is approximately
given by $n_i/N_{2}$ before $\overline{\kappa}_{M1i}$ starts to
differ significantly from $0$.

For the ODEs to be given shortly, the
rapid energy transfer from type--2 particles to
type--1 particles is governed by a coupling constant
$\lambdaTwoOne\geq 0$ and a burst function $\widehat{\omega}(\tau)$,
which is taken to peak around $\tau=\tau_{21}\ll 1$.
For simplicity, define this burst function $\widehat{\omega}$ using the
previous function $\gamma$ from \eqref{eq:function-gamma}
for a fixed value of $\tau_\text{freeze}$:
\bsubeqs\label{eq:functions-AppNew}
\beqa
\widehat{\omega}(\tau)&=&\widehat{\gamma}_{21}(\tau)\,[1-\widehat{\gamma}_{21}(\tau)]\,,
\\[2mm]
\widehat{\gamma}_{21}(\tau) &\equiv& \widehat{\gamma}(\tau/\tau_{21})\,,
\\[2mm]
\widehat{\gamma}(\tau) &\equiv& \gamma(\tau)\,\big|_{\tau_\text{freeze}=3}\;.
\eeqa
The subsequent decay of type--1 particles into type--2
particles is governed by a coupling constant
$\lambdaOneTwo \geq 0$ and a decay function $\widehat{\nu}(\tau)$.
For simplicity, this function is taken to have the following form:
\beq
\widehat{\nu}(\tau) = 1-\widehat{\gamma}(\tau) \,.
\eeq
\esubeqs
The hats on the above functions indicate that they are independent
of the value of the dissipative coupling constant $\zeta$,
whereas $\gamma$ from \eqref{eq:tau_freeze+function-gamma}
does depend on it.
The model functions used will also be shown in the plots
of the numerical results later on.

\subsection{ODEs and  boundary conditions}
\label{subsec:appA-ODEs-and-BCS}

With a type--1 mass spectrum of the
form \eqref{eq:caseAandB-mass-spectrum-AppNew}
and the dynamic EOS function $\overline{\kappa}_{M1i}$
from \eqref{eq:kappaM-wM-alphabeta-theta-AppNew},
the dimensionless ODEs are now taken to be the following:%
\vspace*{2mm}
\bsubeqs\label{eq:modifiedODEs-AppNew-dimensionless}
\beqa
\hspace*{-9mm}
3\,  \big(\dot{h} +2h^2 \big) &=& \gamma\,x
+ (1- \gamma)\,\Big(\xi^{-1}\, x^2
+ \half\,\overline{\kappa}_{M1a}\, r_{M1a}+ \half\,\overline{\kappa}_{M1b}\, r_{M1b}
\Big)\,,
\label{eq:modifiedODEs-AppNew-hdot}
\\[2mm]
\hspace*{-9mm}
\dot{r}_{M1a}+(4-\overline{\kappa}_{M1a})\,h\,r_{M1a}
&=&
(N_{1a}/N_1) \Big(\lambdaTwoOne\,\widehat{\omega}\,r_{M2}-(\zeta/\gamma)\;\big[\dot{x}]\Big)
- \lambdaOneTwo\, \widehat{\nu}\,r_{M1a}  \,,
\label{eq:modifiedODEs-AppNew-rM1adot}
\\[2mm]
\hspace*{-9mm}
\dot{r}_{M1b}+(4-\overline{\kappa}_{M1b})\,h\,r_{M1b}
&=&
(N_{1b}/N_1) \Big(\lambdaTwoOne\,\widehat{\omega}\,r_{M2}-(\zeta/\gamma)\;[\dot{x}]\Big)
- \lambdaOneTwo\, \widehat{\nu}\,r_{M1b}  \,,
\label{eq:modifiedODEs-AppNew-rM1bdot}
\\[2mm]
\hspace*{-11mm}
\dot{r}_{M2}+4\,h\,r_{M2}&=&
-\lambdaTwoOne\,\widehat{\omega}\,r_{M2}+\lambdaOneTwo\, \widehat{\nu}\,(r_{M1a}+r_{M1b}) \,,
\label{eq:modifiedODEs-AppNew-rM2dot}
\\[2mm]
\hspace*{-9mm}
3\,\left( \xi^{-1}\,h\,\dot{x}+ h^2 \right)&=& \xi^{-1}\,x^2/2
+r_{M1a}+r_{M1b}+r_{M2} \,,
\label{eq:modifiedODEs-AppNew-xdot}
\eeqa
\esubeqs
where the dissipation function $\gamma$ is given
by \eqref{eq:tau_freeze+function-gamma} and the other model functions
by \eqref{eq:functions-AppNew}. The basic structure of these ODEs
is identical to that of \eqref{eq:modifiedODEs-dimensionless}
with a prescribed EOS function \eqref{eq:function-kappaM1}.

As the physical hierarchy parameter $\xi$ can be expected to be very large
(perhaps of order $10^{60}$), the $\xi^{-1}=0$ ODEs are especially
relevant:
\bsubeqs\label{eq:modifiedODEs-AppNew-dimensionless-xiinfty}
\beqa
\hspace*{-11mm}
3\,  \big(\dot{h} +2h^2 \big) &=& \gamma\,x
+(1-\gamma)\,
\Big(\half\,\overline{\kappa}_{M1a}\;[3\,  h^2 -r_{M2}-r_{M1b}]+
     \half\,\overline{\kappa}_{M1b}\,r_{M1b}\Big),
\label{eq:modifiedODEs-AppNew-hdot-xiinfty}
\\[2mm]
\hspace*{-11mm}
\dot{x} &=& - \zeta^{-1}\,h\,\gamma^2\,
\Big(2\,x -\overline{\kappa}_{M1a}\;[3\,  h^2 -r_{M2}-r_{M1b}]
          -\overline{\kappa}_{M1b}\,r_{M1b}\,\Big),
\label{eq:modifiedODEs-AppNew-xdot-xiinfty}
\\[2mm]
\hspace*{-11mm}
\dot{r}_{M1b}+(4-\overline{\kappa}_{M1b})\,h\,r_{M1b}
&=&
(N_{1b}/N_1) \Big(\lambdaTwoOne\,\widehat{\omega}\,r_{M2}-(\zeta/\gamma)\;[\dot{x}]\Big)
- \lambdaOneTwo\, \widehat{\nu}\,r_{M1b}  \,,
\label{eq:modifiedODEs-AppNew-rM1bdot-xiinfty}
\\[2mm]
\hspace*{-11mm}
\dot{r}_{M2}+4\,h\,r_{M2}&=&
-\lambdaTwoOne\,\widehat{\omega}\,r_{M2}+
\lambdaOneTwo\, \widehat{\nu}\;[3\,  h^2 -r_{M2}] \,,
\label{eq:modifiedODEs-AppNew-rM2dot-xiinfty}
\eeqa
\esubeqs
with $r_{M1a}$ following from the solution of these ODEs by
the corresponding Friedmann equation,
\beqa
r_{M1a} &=& 3\,  h^2 -r_{M2} -r_{M1b}\,.
\label{eq:modifiedEq-AppNew-Friedmann-xiinfty}
\eeqa
\vspace*{2mm}

For both sets of ODEs, \eqref{eq:modifiedODEs-AppNew-dimensionless}
and \eqref{eq:modifiedODEs-AppNew-dimensionless-xiinfty},
the following boundary conditions on $h$, $r_{M1i}$, $r_{M2}$, and $x$
are taken to
hold at a time $\tau=\tau_\text{min} \ll 1$:
\bsubeqs\label{eq:modifiedODEs-AppNew-dimensionless-bcs}
\beqa
h(\tau_\text{min})&=& 1/2 \;(\tau_\text{min})^{-1} \,,\\[2mm]
r_{M1i}(\tau_\text{min})&=& 0\,,\\[2mm]
r_{M2}(\tau_\text{min})&=& 3 \;\big[h(\tau_\text{min})\big]^2\,,\\[2mm]
x(\tau_\text{min})&=& 0 \,,
\eeqa
\esubeqs
which correspond to a standard radiation-dominated FRW universe.
It must be emphasized that there is essentially no free parameter in
these boundary conditions, the precise value of $\tau_\text{min}$ being
irrelevant as long as it is sufficiently small compared to $1$
(so that $t_\text{min}\ll t_\text{ew}$).

Before turning to the numerical solutions,
it may be helpful to discuss very briefly the heuristics of the
ODEs \eqref{eq:modifiedODEs-AppNew-dimensionless},
the discussion of Eqs.~\eqref{eq:modifiedODEs-AppNew-dimensionless-xiinfty}
and \eqref{eq:modifiedEq-AppNew-Friedmann-xiinfty} being similar.
Starting from the values \eqref{eq:modifiedODEs-AppNew-dimensionless-bcs},
the $\lambdaTwoOne\,\widehat{\omega}$ terms in
\eqref{eq:modifiedODEs-AppNew-rM1adot} and
\eqref{eq:modifiedODEs-AppNew-rM1bdot}
generate a nonzero value of the type--1 matter energy density
$r_{M1a}+r_{M1b}$, which
then gives a nonzero value of the vacuum variable $x$
from \eqref{eq:modifiedODEs-AppNew-xdot}.
The values of $r_{M1i}(\tau)$ and $x(\tau)$ peak around $\tau=1$,
which provides a dynamic version of the electroweak kick,
improving upon the prescribed (artificial) kick discussed
in Ref.~\cite{KV2009-electroweak} and
the main text of the present article.\footnote{It is
clear from \eqref{eq:modifiedODEs-AppNew-xdot-xiinfty}, in particular,
that the kick of the vacuum energy density is driven by
the combination $\sum_{i}\overline{\kappa}_{M1i}\,r_{M1i}$.
\label{ftn:kicks}}
Finally, the value of $x$ approaches a constant
positive value and so does the vacuum energy density $r_V=\half\,x^2$,
whereas $r_{M1a}+r_{M1b}$ drops to zero because of the decay
terms $\lambdaOneTwo\, \widehat{\nu}$ in
\eqref{eq:modifiedODEs-AppNew-rM1adot} and
\eqref{eq:modifiedODEs-AppNew-rM1bdot}.

\subsection{Numerical solution}
\label{subsec:appA-Numerical-solution}
\vspace*{0mm}

\begin{figure*}
\vspace*{-.0cm}
\includegraphics[width=\textwidthPreprintTwocolumn]{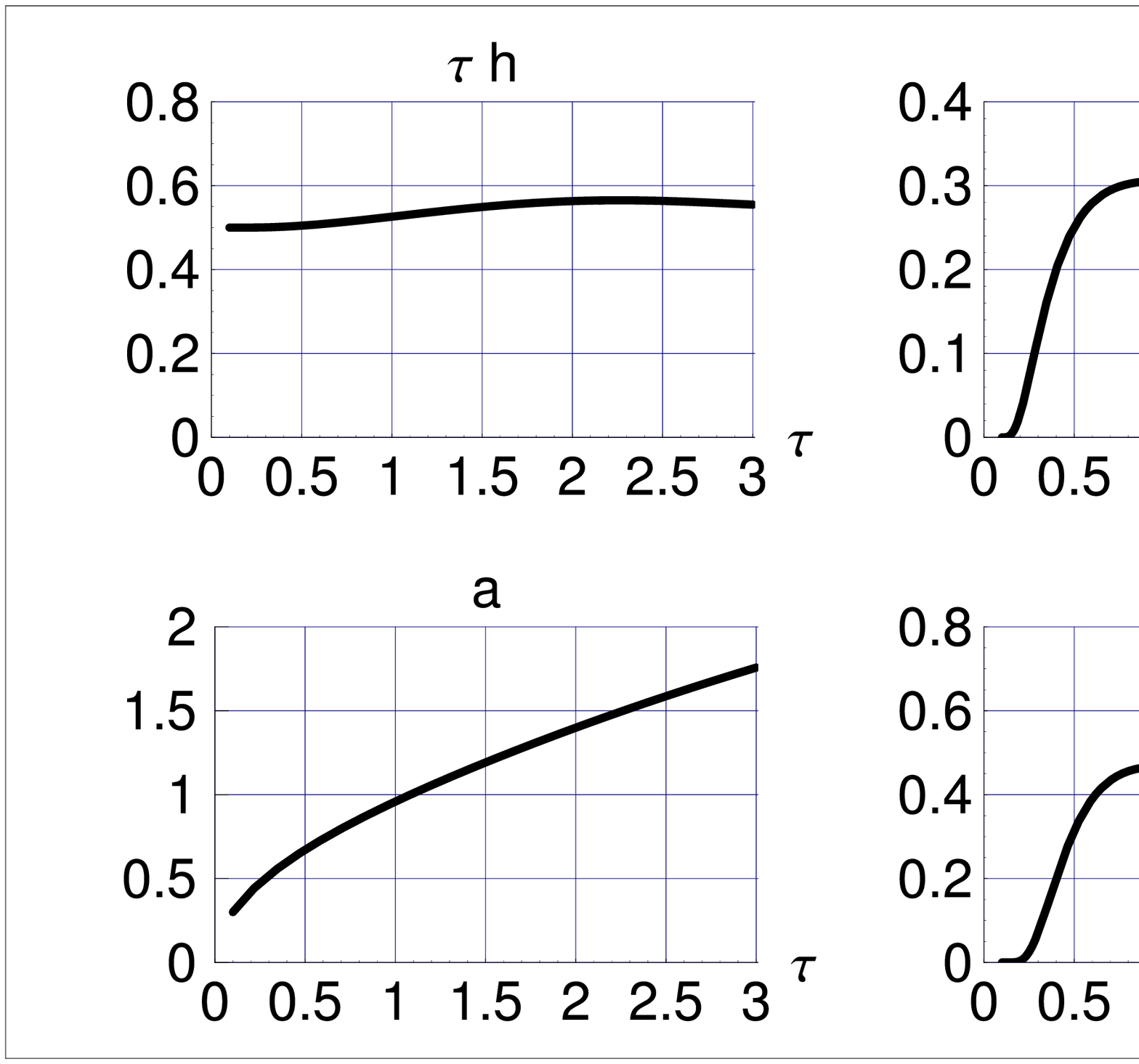} 
\vspace*{3mm}
\caption{Same as Fig.~\ref{fig:1}, but now for nonvanishing dissipative
coupling constant $\zeta=2$ and nontrivial
function $\gamma(\tau)$ defined by \eqref{eq:tau_freeze+function-gamma}
with parameter $\tau_{\infty}=2$, giving $\tau_\text{freeze} = 3$.}
\label{fig:2} \vspace{0mm}
\end{figure*}

\begin{figure*}
\vspace*{-.0cm}
\includegraphics[width=\textwidthPreprintTwocolumn]{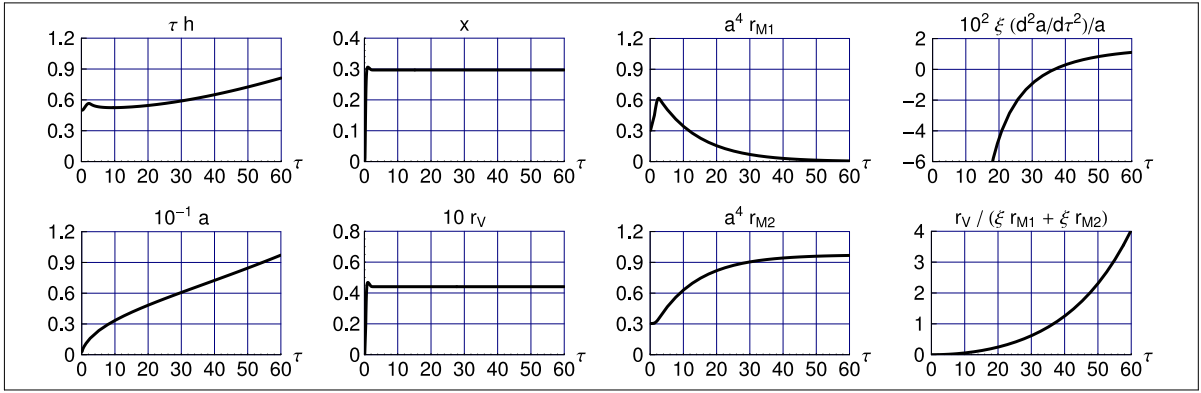}
\vspace*{3mm}
\caption{Same as Fig.~\ref{fig:2}, but also evolved
for $\tau > \tau_\text{freeze} = 3$ with the standard FRW equations
[given by \eqref{eq:modifiedODEs-dimensionless} for $\dot{x}=0$,
$\kappa_{M1}=0$, and $\gamma=0$].}
\label{fig:3} \vspace{0mm}
\end{figure*}

\begin{figure*}
\vspace*{0cm}
\includegraphics[width=\textwidthPreprintTwocolumn]{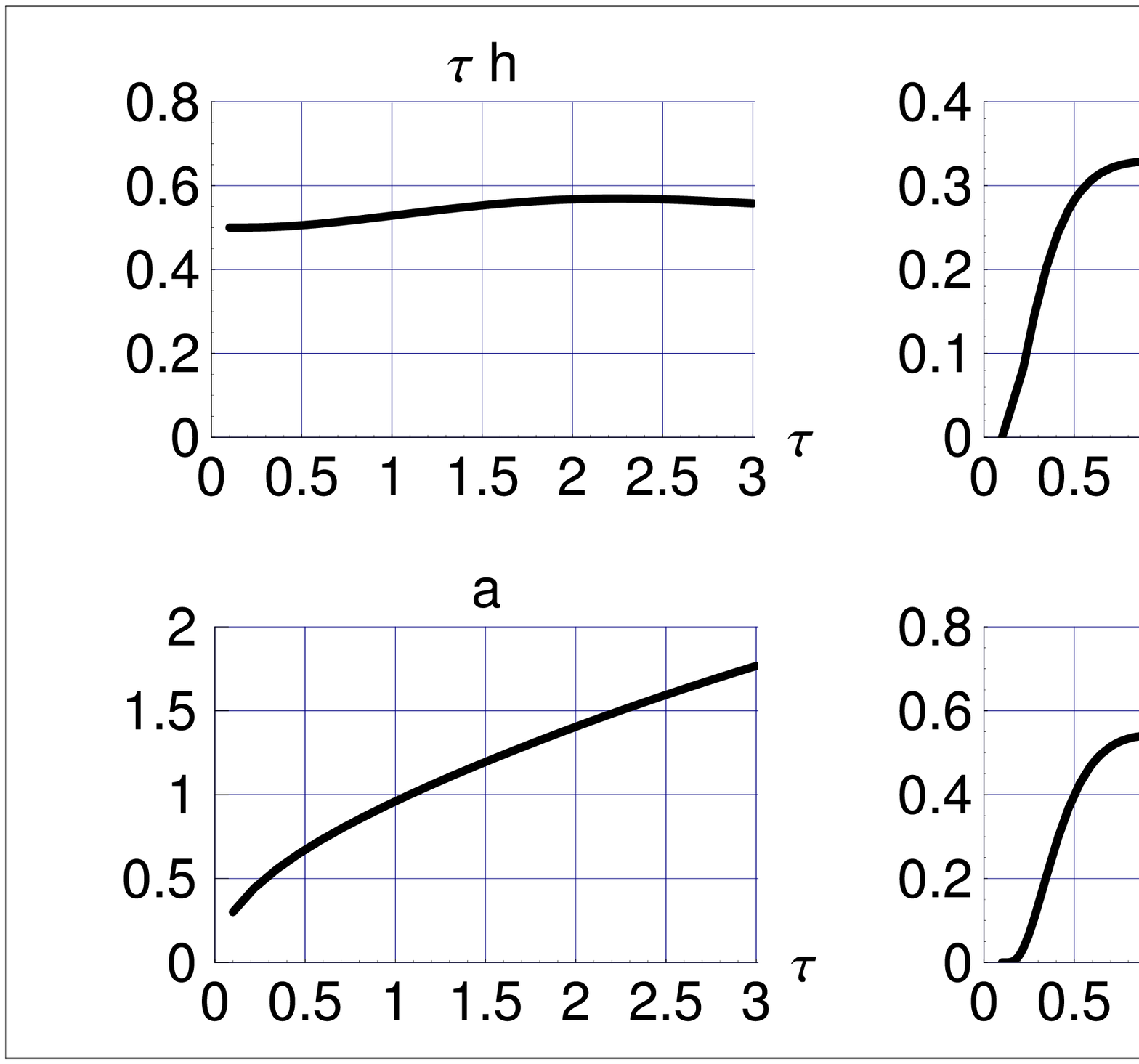}
\vspace*{3mm}
\caption{Same as Fig.~\ref{fig:2}, still with $\zeta=2$
but now for a larger hierarchy parameter $\xi=10^{4}$.}
\label{fig:5} \vspace{0mm}
\end{figure*}

\begin{figure*}
\vspace*{-.0cm}
\includegraphics[width=\textwidthPreprintTwocolumn]{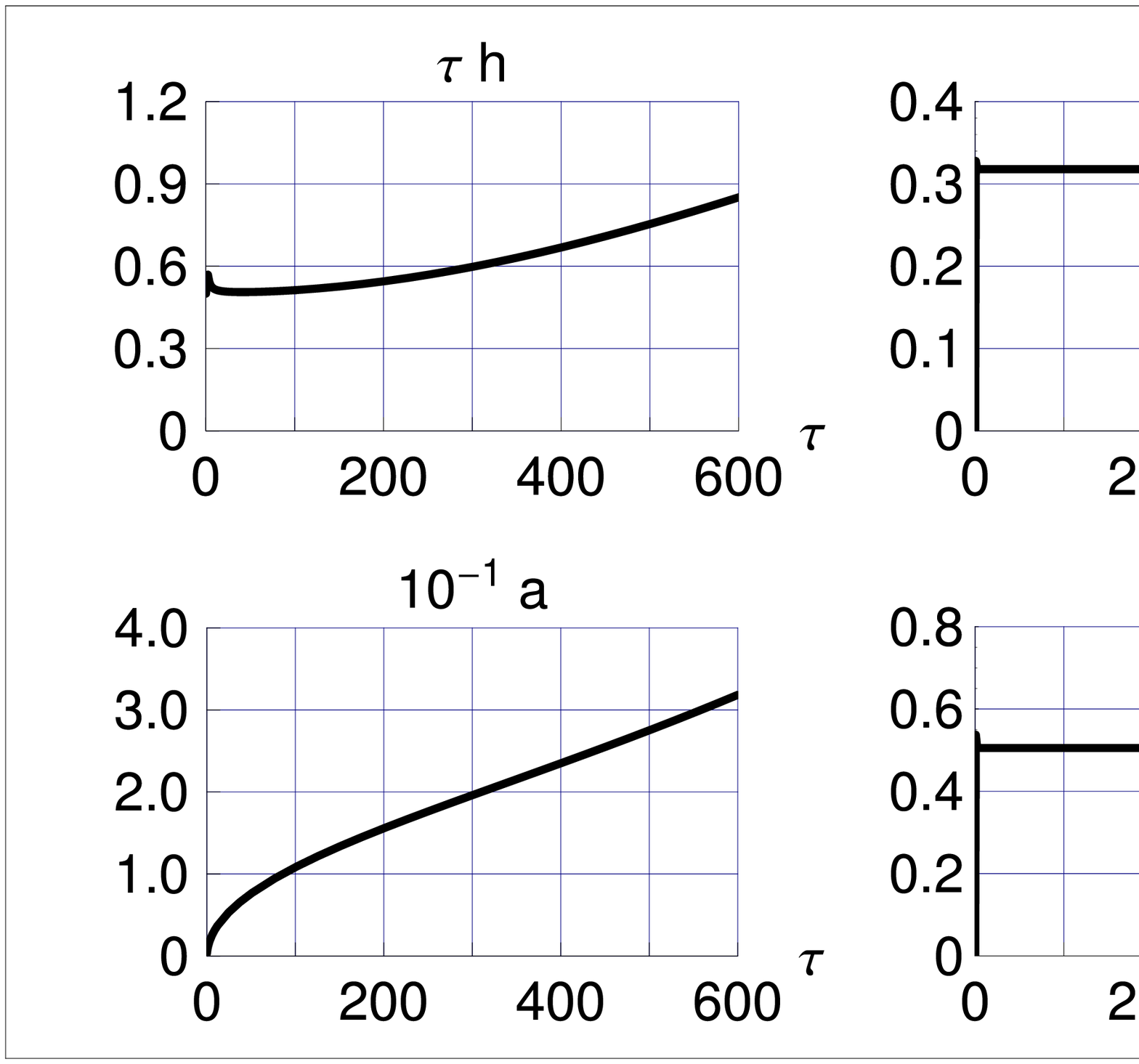}
\vspace*{3mm}
\caption{Same as Fig.~\ref{fig:3}, still with $\zeta=2$
but now for a larger hierarchy parameter $\xi=10^{4}$.}
\label{fig:6} \vspace*{0cm}
\end{figure*}

\begin{figure*}   [t]     
\vspace*{-.0cm}
\includegraphics[width=0.85\textwidth]{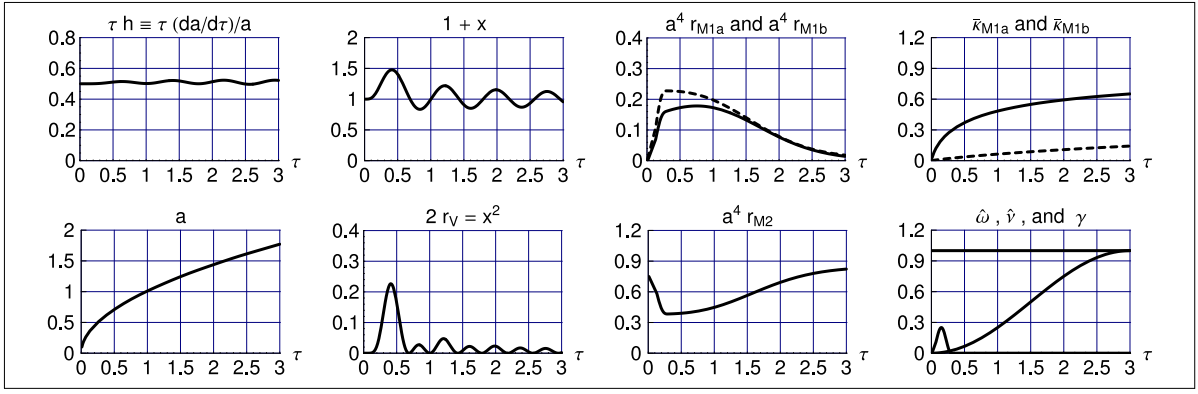}
\vspace*{3mm}
\caption{Numerical solution of the ODEs \eqref{eq:modifiedODEs-AppNew-dimensionless}
with the dynamic EOS function $\overline{\kappa}_{M1}$ from
\eqref{eq:kappaM-wM-alphabeta-theta-AppNew}, for vanishing  dissipative
coupling constant $\zeta=0$ and trivial function $\gamma(\tau)=1$.
Model parameters are $\{\xi,\,      \lambdaTwoOne  ,\,  \lambdaOneTwo \}$=
                             $\{100,\,   18  ,\,             2 \}$
and model functions are defined by \eqref{eq:functions-AppNew}
with parameter $\tau_{21}=1/10$
[the function $\widehat{\omega}(\tau)$ peaks below $\tau=0.3$
and $\widehat{\nu}(\tau)$ rises from $0$ to $1$].
The case--A type--1 mass spectrum \eqref{eq:caseA-mass-spectrum-AppNew}
has been assumed,
with type--1a functions shown by full curves and type--1b functions by dashed curves.
The ODEs are solved over the interval
$[\tau_\text{min},\, \tau_\text{max}]$ $=$ $[0.01,\,  3]$ with the following
boundary conditions from \eqref{eq:modifiedODEs-AppNew-dimensionless-bcs}
at $\tau=\tau_\text{min}=0.01$:
$\{x,\, h,\,   a,\,  r_{M1a},\,r_{M1b},\, r_{M2}\}$ $=$
$\{0,\, 50,\,  0.1,\,0,\,      0,\,       7500 \}$.
}
\label{fig:8} \vspace*{0cm}
\end{figure*}

\begin{figure*}
\vspace*{-.0cm}
\includegraphics[width=0.85\textwidth]{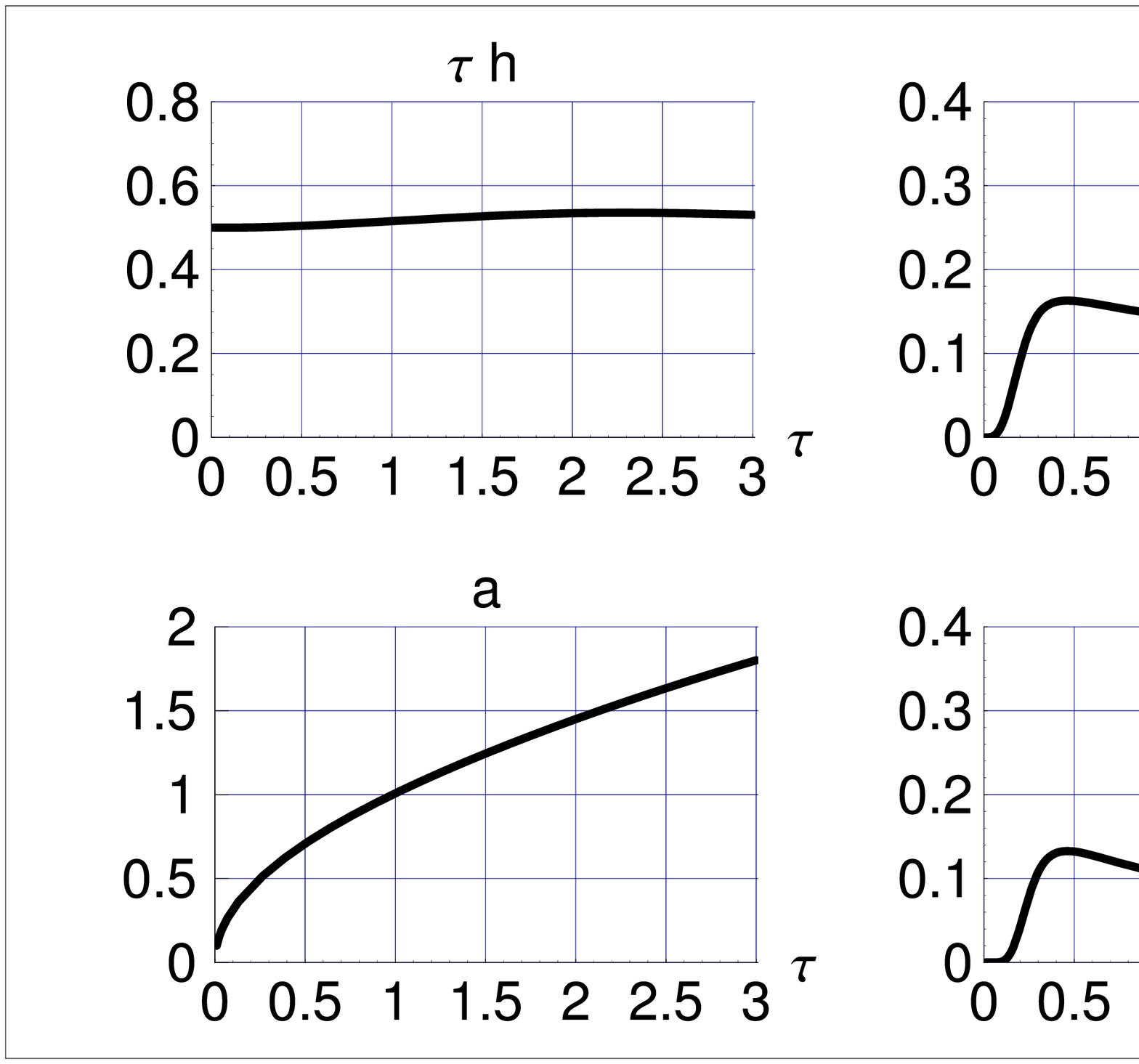}
\vspace*{3mm}
\caption{Same as Fig.~\ref{fig:8}, but now for nonvanishing
dissipative coupling constant $\zeta=2$ and nontrivial
function $\gamma(\tau)$ defined by \eqref{eq:tau_freeze+function-gamma}
with parameter $\tau_{\infty}=2$, giving $\tau_\text{freeze} = 3$.}
\label{fig:9} \vspace*{0mm}
\end{figure*}

\begin{figure*}
\vspace*{0cm}
\includegraphics[width=0.85\textwidth]{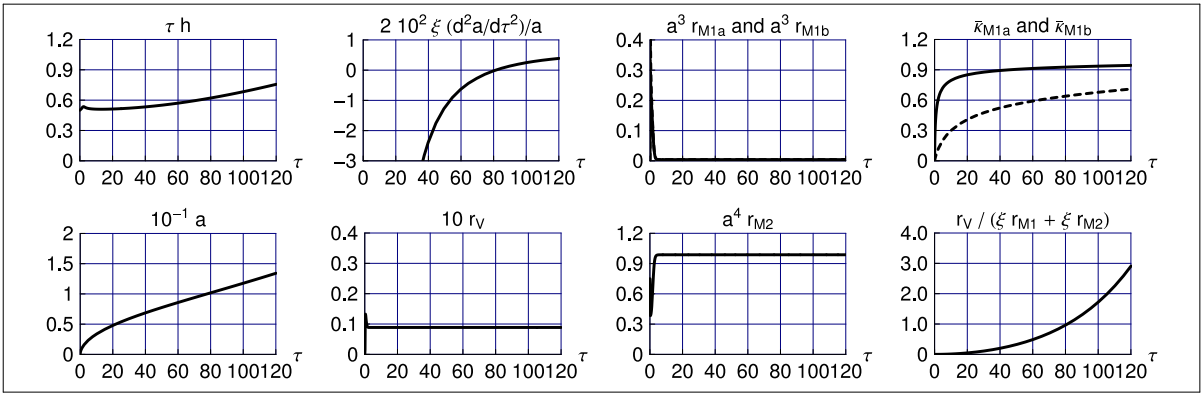}
\vspace*{3mm}
\caption{Same as Fig.~\ref{fig:9}, but also evolved
for $\tau > \tau_\text{freeze} = 3$.}
\label{fig:10}
\vspace*{0mm}
\end{figure*}

Numerical results for the case--A mass
spectrum \eqref{eq:caseA-mass-spectrum-AppNew}
and hierarchy parameter $\xi=10^2$
are given in Fig.~\ref{fig:8} for the standard (nondissipative)
$q$--theory ODEs with $\zeta=0$ and
in Figs.~\ref{fig:9}--\ref{fig:10} for
the modified (dissipative) $q$--theory ODEs with $\zeta=2$.
[The coupling constant value $\lambdaTwoOne=18$ has been chosen
to give $r_{M1a}+r_{M1b} \sim r_{M2}$ at $\tau=0.25$,
as would approximately correspond to thermal equilibrium.
The value of $\lambdaOneTwo=2$ has been chosen so that most
type--1 particles have decayed by a cosmic time $\tau=3$.
But the results for $r_V^\text{num}$ are more or
less constant for $\lambdaOneTwo\lesssim 20$, which includes
the nonrealistic case $\lambdaOneTwo=0$ of stable (massive)
type--1 particles.]
Figures~\ref{fig:9} and \ref{fig:10} are seen to give
a somewhat lower value of the dimensionless remnant vacuum energy density
$r_V^\text{num}\equiv r_{V}(\tau_\text{freeze})$
than Figs.~\ref{fig:2} and \ref{fig:3},
one reason being that the maximum value of $\overline{\kappa}_{M1}\,r_{M1}$
is less for the dynamic kick considered than for
the prescribed kick (cf. Ftn.~\ref{ftn:kicks}).

\begin{figure*}[t]
\vspace*{-.0cm}
\includegraphics[width=0.85\textwidth]{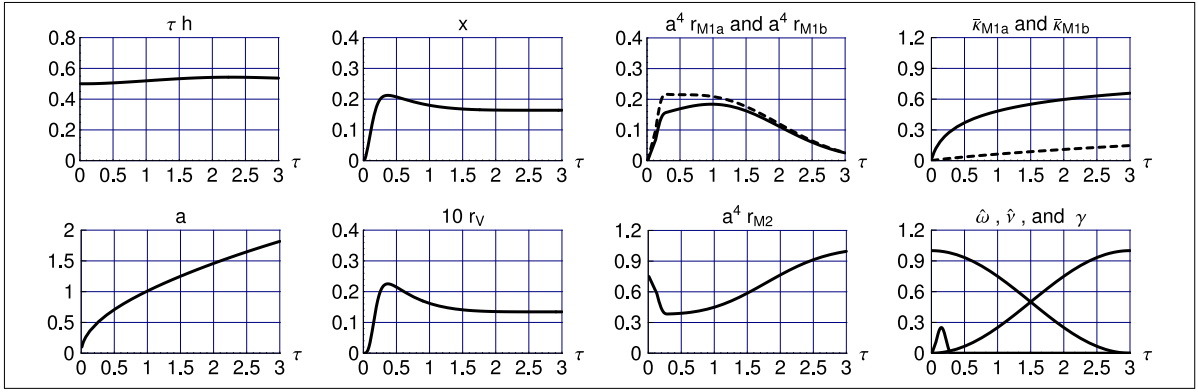}
\vspace*{3mm}
\caption{Same as Fig.~\ref{fig:9}, still for the case--A mass
spectrum \eqref{eq:caseA-mass-spectrum-AppNew} but now solving the
$\xi^{-1}=0$ ODEs \eqref{eq:modifiedODEs-AppNew-dimensionless-xiinfty}
and using the algebraic Eq.~\eqref{eq:modifiedEq-AppNew-Friedmann-xiinfty}.
}
\label{fig:11} \vspace*{0mm}
\end{figure*}

\begin{figure*}   
\vspace*{0mm}
\includegraphics[width=0.85\textwidth]{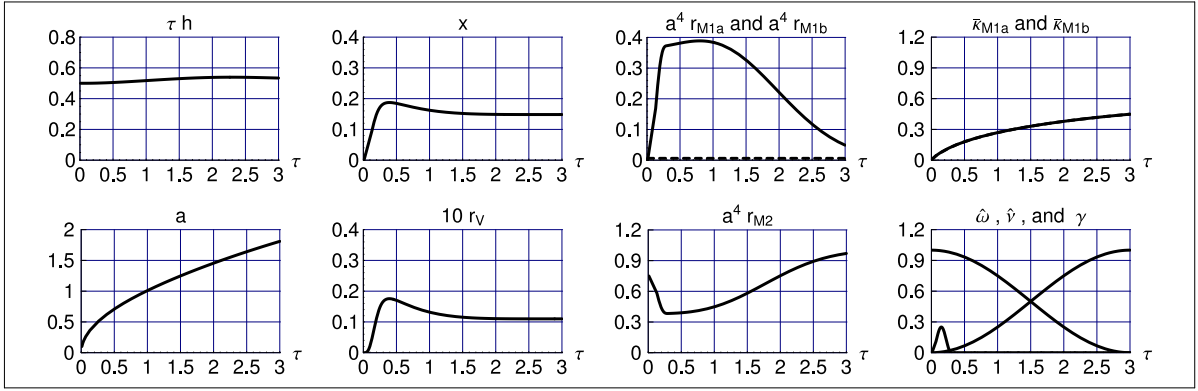}
\vspace*{3mm}
\caption{Same as Fig.~\ref{fig:11}, still with $\xi^{-1}=0$ but now
for the case--B mass spectrum \eqref{eq:caseB-mass-spectrum-AppNew}.}
\label{fig:12} \vspace*{10mm}
\vspace*{0mm}
\end{figure*}

Numerical results for
the case--A mass spectrum \eqref{eq:caseA-mass-spectrum-AppNew} and
the physically relevant
hierarchy parameter $\xi=\infty$ are given in Fig.~\ref{fig:11},
with $r_V^\text{num}$ of order $1.34 \times 10^{-2}$
for $\zeta=2$.
Similar results (Fig.~\ref{fig:12}) are obtained
for the case--B mass spectrum \eqref{eq:caseB-mass-spectrum-AppNew},
with $r_V^\text{num}$ of order $1.10 \times 10^{-2}$.
As mentioned in Sec.~\ref{sec:Introduction},
Figs.~\ref{fig:11} and \ref{fig:12} present the best (i.e., most
realistic) calculations of this article.

According to \eqref{eq:Eew-estimate}, an average value
$r_V^\text{num}\sim 1.2 \times 10^{-2}$
has a required $E_\text{ew}$ value of order $3.8\;\text{TeV}$,
which is 20\% above the value
$3.2\;\text{TeV}$ from Table~\ref{tab-Neff1-Eew}.
The mass values corresponding to $E_\text{ew}\sim 3.8\;\text{TeV}$ are
$M_{1a}\sim 7.7\;\text{TeV}$ and $M_{1b}\sim 1.3\;\text{TeV}$
for the case--A spectrum and $M_{1a}\sim 3.8\;\text{TeV}$ for
the case--B spectrum.
With the dynamic kick as defined in this appendix,
the  obtained value $E_\text{ew}\sim 3.8\;\text{TeV}$
depends primarily on the assumption that
the dissipation function $\gamma$ enters the
ODEs \eqref{eq:modifiedODEs-AppNew-dimensionless-xiinfty} in the way
shown and that the relevant coupling constant $\zeta$
is of order unity.

Consider, then, the $\zeta$ dependence
of the predicted $E_\text{ew}$ value. As explained
in the fourth paragraph of Sec.~\ref{subsec:Additional-remarks},
the value of the remnant vacuum energy density $r_V^\text{num}(\zeta)$
can be expected to drop to zero for $\zeta \downarrow 0$
and $\zeta\to\infty$. From the numerical solution
of ODEs \eqref{eq:modifiedODEs-AppNew-dimensionless-xiinfty}
for the case--A mass spectrum \eqref{eq:caseA-mass-spectrum-AppNew}
and the same boundary conditions as in Fig.~\ref{fig:11},
the following rough fit of $r_V^\text{num}(\zeta)$ for $\zeta\geq 0.02$
is found: $r_V^\text{fit}(\zeta)=
\zeta^5\big/\big(\,\widetilde{\alpha}\, \zeta^7+\widetilde{\beta}\,\big)$
with constants $(\widetilde{\alpha},\,\widetilde{\beta})=
(13.29,\, 680.3)$.
More precisely, the $r_V^\text{num}(\zeta)$ value peaks
at $1.34\times 10^{-2}$ for $\zeta=2$
and drops to $2.46 \times 10^{-7}$ for $\zeta=0.2$
and to $5.82 \times 10^{-4}$ for $\zeta=20$.
[Note that, for $\zeta \ll 0.2$, the time scale
$\tau_\text{freeze}$  from \eqref{eq:tau_freeze}
may become unreasonably large and that, for $\zeta \gg 20$,
the particle-production effects on the right-hand side
of \eqref{eq:modifiedODEs-AppNew-rM1bdot-xiinfty},
for given value of $\dot{x} \propto \dot{\rho}_V$,
may be larger than can be expected from weakly-coupled particles.]
The corresponding results for $E_\text{ew}$ are given in
Table~\ref{tab-zeta-Eew}, which, together with Table~\ref{tab-Neff1-Eew},
gives an idea of the systematic uncertainties involved
(compare the $E_\text{ew}$ values for
$N_{\text{eff},1}=10^2$ and $\zeta=2$ from both tables).

For the dynamic kick considered in this Appendix, the estimate is thus
\beq\label{eq:Eew-AppNew}
E_\text{ew}\;
\big|^\text{\,dynamic kick}_
{N_{\text{eff},1}=N_{\text{eff},2}=10^2,\;\text{case}-A}
\sim 4 - 15\;\text{TeV}\,,
\eeq
assuming that the dissipation function $\gamma$ enters the
ODEs \eqref{eq:modifiedODEs-AppNew-dimensionless-xiinfty} in the way
shown and that the coupling constant $\zeta$ lies between 0.2 and 20.
The corresponding case--A mass values are $M_{1a}=2\,E_\text{ew}$ and
$M_{1b}=1/3\,E_\text{ew}$.
The case--B mass spectrum \eqref{eq:caseB-mass-spectrum-AppNew},
with a unique mass value $M_{1a}=E_\text{ew}$,
gives the same $E_\text{ew}$ estimate as in \eqref{eq:Eew-AppNew}.

\end{appendix}


\newpage
\begin{table*}[t]
\begin{center}
\caption{Function values of $h(\tau)$, $r_{M1}(\tau)$, $r_{M2}(\tau)$,
and $r_{V}(\tau)$ at $\tau=\tau_\text{freeze}=3$
from the numerical solution of the ODEs \eqref{eq:modifiedODEs-dimensionless}
for dissipative coupling constant $\zeta=2$, hierarchy parameter
$\xi=10^2$, and various values of
the boundary condition $r_{M1}/r_{M2}$ at $\tau=\tau_\text{min}=0.1$.
The other parameters and boundary conditions
are given in the caption of Fig.~\ref{fig:1}.
The numerical accuracy of the quoted function values is estimated
to be equal to $\pm 1$ in the last digit.\vspace*{.5\baselineskip}}
\label{tab-rM1rM2-xi100}
\renewcommand{\tabcolsep}{1.5pc}    
\renewcommand{\arraystretch}{1.1}   
\begin{tabular}{cc|cccc}
\hline\hline
$\xi$  & $\big[r_{M1}/r_{M2}\big](\tau_\text{min})$  &
$h$ & $r_{M1}$  &$r_{M2}$ & $r_{V}$\\
\hline
$10^{2}$ &$10^{-1}$            &  $0.1702$& $0.01329$&            $0.07357$& $0.001473$\\
$10^{2}$ &$1$                  &  $0.1849$& $0.06321$&            $0.03893$& $0.04407\phantom{0}$\\
$10^{2}$ &$10^{1}$ &  $0.1983$& $0.1007\phantom{0}$&  $0.01577$& $0.1441\phantom{00}$\\
$10^{2}$ &$10^{2}$ &  $0.2008$& $0.1070\phantom{0}$&  $0.01220$& $0.1705\phantom{00}$\\
\hline\hline
\end{tabular}
\end{center}
\vspace*{0mm}
\end{table*}

\begin{table*}
\begin{center}
\caption{Same as Table~\ref{tab-rM1rM2-xi100}, but now for a fixed
boundary condition $[r_{M1}/r_{M2}](t_\text{min})=1$ and
various hierarchy parameters $\xi$ ranging from $1$ to $10^6$.
The entry for $\xi=\infty$ has been calculated from the ODEs
\eqref{eq:modifiedODEs-dimensionless-xiinfty} and the
algebraic equation \eqref{eq:modifiedEq-Friedmann-xiinfty}.
\vspace*{.5\baselineskip}}
\label{tab-xi-extrapol}
\renewcommand{\tabcolsep}{1.5pc}    
\renewcommand{\arraystretch}{1.1}   
\begin{tabular}{cc|cccc}
\hline\hline
 $\xi$ & $\big[r_{M1}/r_{M2}\big](\tau_\text{min})$  &
 $h$ & $r_{M1}$  &$r_{M2}$ & $r_{V}$\\
\hline
$1$       & $1$ & $0.1758$ & $0.04279$ & $0.04394$ & $0.006026$\\
 $10^{1}$ & $1$ & $0.1817$ & $0.05582$ & $0.04069$ & $0.02584\phantom{0}$\\
 $10^{2}$ & $1$ & $0.1849$ & $0.06321$ & $0.03893$ & $0.04407\phantom{0}$\\
 $10^{4}$ & $1$ & $0.1859$ & $0.06530$ & $0.03841$ & $0.05052\phantom{0}$\\
 $10^{6}$ & $1$ & $0.1859$ & $0.06533$ & $0.03841$ & $0.05061\phantom{0}$\\
 $\infty$ & $1$ & $0.1860$ & $0.06533$ & $0.03841$ & $0.05061\phantom{0}$\\
\hline\hline
\end{tabular}
\end{center}
\vspace*{0mm}
\end{table*}

\begin{table*}
\begin{center}
\caption{Same as Table~\ref{tab-rM1rM2-xi100}, but now from
the $\xi^{-1}=0$ ODEs \eqref{eq:modifiedODEs-dimensionless-xiinfty}
and Eq.~\eqref{eq:modifiedEq-Friedmann-xiinfty}.%
\vspace*{0.25\baselineskip}}
\label{tab-rM1rM2-xiinfty}
\renewcommand{\tabcolsep}{1.25pc}    
\renewcommand{\arraystretch}{1.1}   
\begin{tabular}{cc|cccc}
\hline\hline
 $\xi$  &  $\big[r_{M1}/r_{M2}\big](\tau_\text{min})$ &
 $h$ & $r_{M1}$  &$r_{M2}$ & $r_{V}$\\
\hline
$\infty$ & $10^{-3} $          & $0.1667$ & $0.0001577$            & $0.08322$ & $2.058 \times 10^{-7}$\\
$\infty$ & $10^{-2}$           & $0.1671$ & $0.001557 \phantom{0}$ & $0.08219$ & $2.021 \times 10^{-5}$\\
$\infty$ & $10^{-1}$           & $0.1704$ & $0.01384 \phantom{00}$ & $0.07330$ & $0.001699$\\
$\infty$ & $\phantom{0}1^{\phantom{-1}}$
& $0.1860$ & $0.06533 \phantom{00}$ & $0.03841$ & $0.05061 \phantom{0}$\\
$\infty$ & $10^{1 \phantom{-}}$ & $0.1995$ & $0.1036 \phantom{000}$ & $0.01583$ & $0.1647 \phantom{00}$\\
$\infty$ & $10^{2 \phantom{-}}$ & $0.2020$ & $0.1100 \phantom{000}$ & $0.01243$ & $0.1948 \phantom{00}$\\
$\infty$ & $10^{3 \phantom{-}}$ & $0.2023$ &  $0.1107 \phantom{000}$ &  $0.01207$ &  $0.1983 \phantom{00}$\\
\hline\hline
\end{tabular}
\end{center}
\end{table*}

\newpage
\begin{table*}
\vspace*{-.5cm}
\begin{center}
\caption{Estimates for the energy scale $E_\text{ew}$
of type--1 particles with mass scale $M \sim E_\text{ew}$
as a function of their effective number of degrees of freedom
$N_{\text{eff},1}$. The inverse of the hierarchy parameter
\eqref{eq:xi-def} is taken to vanish, $\xi^{-1}=0$,
and the dissipative coupling constant is assumed to have a value of
order unity, specifically $\zeta=2$.
The kick from the type--1 particles is modeled by the prescribed
EOS function $\kappa_{M1}(\tau)$ from \eqref{eq:function-kappaM1}
with parameters listed in the caption of Fig.~\ref{fig:1}.
Both ultramassive type--1 and massless type--2 particles
are assumed to have been in thermal equilibrium for $T\gg E_\text{ew}$
and the effective number of type--2 particles is taken as
$N_{\text{eff},2}=10^2$.  The $E_\text{ew}$ estimates are obtained
from \eqref{eq:Eew-estimate}, using the $r_V$ values from
Table~\ref{tab-rM1rM2-xiinfty}
for $[r_{M1}/r_{M2}\big](\tau_\text{min})=N_{\text{eff},1}/10^2$
and taking the measured value of the
cosmological constant to be $\Lambda^\text{exp} = (2\;\text{meV})^4$.
\vspace*{0.25\baselineskip}}
\label{tab-Neff1-Eew}
\renewcommand{\tabcolsep}{1.5pc}    
\renewcommand{\arraystretch}{1.1}   
\begin{tabular}{cc|c}
\hline\hline
 $\zeta$ & $N_{\text{eff},1}$ &
 $E_\text{ew}^\text{\,prescribed kick}\,[\text{TeV}]$ \\
\hline
 $2$ &$1    $ & $8.5$ \\   
 $2$ &$10^1 $ & $4.9$ \\ 
 $2$ &$10^2 $ & $3.2$ \\  
 $2$ &$10^3 $ & $2.8$ \\ 
 $2$ &$10^4 $ & $2.7$ \\ 
\hline\hline
\end{tabular}
\end{center}
\end{table*}

\begin{table*}
\vspace*{-1cm}  
\begin{center}
\caption{Same as Table~\ref{tab-Neff1-Eew} but now for the dynamic
EOS function $\overline{\kappa}_{M1}(\tau)$
of Appendix~\ref{sec:appendixA}. The total effective
numbers of degrees of freedom are $N_{\text{eff},1}=N_{\text{eff},2}=10^2$
and the type--1 mass spectrum is given by \eqref{eq:caseA-mass-spectrum-AppNew}.
The estimates for the energy scale $E_\text{ew}$ are given for $\xi^{-1}=0$
and three values of the dissipative
coupling constant $\zeta$, the other
coupling constants being listed in the caption of Fig.~\ref{fig:9}.
\vspace*{0\baselineskip}}
\label{tab-zeta-Eew}
\renewcommand{\tabcolsep}{1.5pc}    
\renewcommand{\arraystretch}{1.1}   
\begin{tabular}{cc|c}
\hline\hline
$\zeta$ & $N_{\text{eff},1}$ &
$E_\text{ew}^\text{\,dynamic kick}\,[\text{TeV}]$ \\
\hline
$0.2   $ &  $10^2 $ & $          14.8$ \\   
$2     $ &  $10^2 $ & $\phantom{0}3.8$ \\   
$20    $ &  $10^2 $ & $\phantom{0}5.6$ \\   
\hline\hline
\end{tabular}
\end{center}
\vspace*{0cm}
\end{table*}





\newpage


\newpage

\newpage

\newpage

\end{document}